\documentclass[journal,twoside]{IEEEtran}
\makeatletter
\long\def\@makecaption#1#2{\ifx\@captype\@IEEEtablestring%
\footnotesize\begin{center}{\normalfont\footnotesize #1}\\
{\normalfont\footnotesize\scshape #2}\end{center}%
\@IEEEtablecaptionsepspace
\else
\@IEEEfigurecaptionsepspace
\setbox\@tempboxa\hbox{\normalfont\footnotesize {#1.}~~ #2}%
\ifdim \wd\@tempboxa >\hsize%
\setbox\@tempboxa\hbox{\normalfont\footnotesize {#1.}~~ }%
\parbox[t]{\hsize}{\normalfont\footnotesize \noindent\unhbox\@tempboxa#2}%
\else
\hbox to\hsize{\normalfont\footnotesize\hfil\box\@tempboxa\hfil}\fi\fi}
\makeatother
\IEEEoverridecommandlockouts
\usepackage{amsthm, fancyhdr, amsmath,amssymb,amsfonts,bm}
\usepackage{array, booktabs}
\usepackage{tipa}
\usepackage{gensymb}
\usepackage{algorithmic}
\usepackage{graphicx}
\usepackage{textcomp}
\usepackage{xcolor}
\usepackage[linesnumbered,ruled]{algorithm2e}

\newtheorem{proposition}{Proposition}

\DeclareMathOperator*{\argmin}{argmin}
\newcommand*{\argminl}{\argmin\limits}
\DeclareMathOperator*{\argmax}{argmax}
\newcommand*{\argmaxl}{\argmax\limits}
\usepackage{dirtytalk}
\ifCLASSOPTIONcompsoc
\usepackage{caption}
\usepackage[caption=false,font=normalsize,labelfon
t=sf,textfont=sf]{subfig}
\else
\usepackage[caption=false,font=footnotesize]{subfi
g}
\fi
\newcommand{\norm}[1]{\left\lVert #1 \right\rVert}
\ifCLASSINFOpdf
 
\else
 
\fi

\begin{document}
%
\title{Graph Laplacian Diffusion Localization of Connected and Automated Vehicles
\thanks{This paper has received funding from the European Union’s
H2020 research and innovation programme CPSoSaware under grant agreement No 871738.}}
%
%
%

\author{Nikos~Piperigkos,
        Aris~S. Lalos,~\IEEEmembership{Senior Member,~IEEE,}
        and~Kostas~Berberidis,~\IEEEmembership{Senior Member,~IEEE}
\thanks{Nikos Piperigkos is with Computer Engineering and Informatics Department, University of Patras, Greece and the Industrial Systems Institute, ATHENA Research Center, Patras, Greece (e-mail: piperigkos@ceid.upatras.gr).}
\thanks{Aris S. Lalos is with Industrial Systems Institute, ATHENA Research Center, Patras, Greece and Electrical and Computer Engineering Department, University of Patras, Greece (e-mail: lalos@isi.gr).}
\thanks{Kostas Berberidis is with Computer Engineering and Informatics Department, University of Patras, Greece (e-mail: berberid@ceid.upatras.gr).}
}

%
%

\markboth{IEEE TRANSACTIONS ON INTELLIGENT TRANSPORTATION SYSTEMS,~Vol.~x, No.~x, Month~2021}
{Piperigkos \MakeLowercase{\textit{et al.}}: Graph Laplacian Diffusion Localization of Connected and Automated Vehicles}

%



\maketitle

\begin{abstract}

In this paper, we design distributed multi-modal localization approaches for Connected and Automated vehicles. We utilize information diffusion on graphs formed by moving vehicles, based on Adapt-then-Combine strategies combined with the Least-Mean-Squares and the Conjugate Gradient algorithms. We treat the vehicular network as an undirected graph, where vehicles communicate with each other by means of Vehicle-to-Vehicle communication protocols. Connected vehicles perform cooperative fusion of different measurement modalities, including location and range measurements, in order to estimate both their positions and the positions of all other networked vehicles, by interacting only with their local neighborhood. The trajectories of vehicles were generated either by a well-known kinematic model, or by using the CARLA autonomous driving simulator. The various proposed distributed and diffusion localization schemes significantly reduce the GPS error and do not only converge to the global solution, but they even outperformed it. Extensive simulation studies highlight the benefits of the various approaches, outperforming the accuracy of the state of the art approaches. The impact of the network connections  and the network  latency  are  also  investigated. 
\end{abstract}


%
\IEEEpeerreviewmaketitle

\section{Introduction}
%
%
%
%
\IEEEPARstart{T}{he} development and growth which has been achieved towards autonomous driving the past few years, will contribute to the reduction of road crashes, improve traffic congestion and fuel consumption and enhance the overall performance of transportation sector. Scientific community and automotive industry is working hard to myriads of applications, such as object detection and tracking, automated parking, traffic signs detection, path planning, cooperative path planning etc. Integrated sensors like GPS, LIDAR, or Camera, to name a few, provide increased perception and scene analysis ability, improving safety. Moreover, Vehicle-to-Everything (V2X) wireless communication technologies among vehicles or Road Side Units (RSU), i.e. Vehicle-to-Vehicle (V2V) and Vehicle-to-Infrastructure (V2I), as well as the prototyping of 5G related protocols, enable smart vehicles of a Vehicular ad-hoc Network (VANET) to have access to external and off-board information. Thus, Connected and Automated Vehicle (CAV) emerges as the type of vehicle of the future, which provides safety, comfort and efficiency to related challenges, facilitating the so-called 4D cooperative awareness (e.g. accurate and timely identification/detection of occluded/non-occluded pedestrian, cyclists and/or vehicles).

Collaborating CAVs usually exploit heterogeneous data and extracted information from sensors, such as absolute position, relative distance, relative angle, velocity, acceleration, etc., to achieve their collective or individual tasks. To this end, Localization, i.e. accurate knowledge of self and other's road users/vehicles location, is considered vital. Although GPS sensor is the most common sensor providing absolute position, it's accuracy is highly degraded in dense urban canyons or tunnels \cite{Kuutti2018}, even exceeding 10$m$ error, while it is also vulnerable to cyber-attacks \cite{Ioannides2016}. Several approaches have been developed for outperforming GPS technology, relying on ground base stations, which are able to achieve localization error in cm level of accuracy. However, they are also vulnerable to multi-path effect and signal blockage \cite{Alam2013}. The desired localization error should be lower than 1$m$ (where-in-lane accuracy) \cite{Neto2020} to meet the standards of autonomous driving. For example, if a vehicle is localized on the curb instead of the road, it may lead to a serious accident with pedestrians or other vehicles. Therefore, assisted or cooperative localization (CL) has received increasing interest as a viable alternative to cope with GPS limitations and erroneous measurements. 

Estimating accurately the location of the ego and also neighboring vehicles, facilitates several other autonomous driving functionalities  including Path Planning, Control tasks and platooning \cite{Montanaro2018}. Path Planning and Control module is responsible for determining and performing the best possible driving actions, e.g. estimate the proper velocity for maintaining safety distances among the vehicles, improving safety and enhancing transportation efficiency. However, distributed CL approaches \cite{Elazab2017}, \cite{Soatti2018}, \cite{Rohani2015} are mainly used only for estimating ego-vehicle's location, by utilizing the local neighborhoods. Moreover, they actually don't evaluate how accurate the ego-vehicle computes the position of neighbors. At the same time, diffusion approaches on graphs \cite{Cattivelli2010a}, \cite{Nassif2020} provide the ability to estimate the locations of connected neighbors. To the best of our knowledge, there is no study up till now about distributed and diffusion vehicular localization, i.e. each vehicle relying only to its neighbors to be in place to estimate the entire common location vector of the VANET which belongs. Our previous work on non-Bayesian \textbf{Centralized Laplacian Localization (CLL)} \cite{Piperigkos2020}, \cite{Piperigkos2020a}, \cite{Piperigkos2020b}, which linearly correlates via the graph Laplacian operator the multi-modal measurement modalities along with the connectivity representation of vehicles, facilitated the design of the proposed approaches. Therefore, in this work we focus on proposing novel diffusion localization schemes for CAVs, which aim to converge to \textbf{CLL} utilizing  the  graph  Laplacian  operator. Note that the proposed approaches can be easily extend to other relevant multi-agent cooperating localization scenarios, like a fleet of robots or Unmanned Aerial Vehicles.  
In summary, the main contributions of our work are:
\begin{itemize}
    \item We propose novel distributed multi-modal localization approaches for CAVs, utilising Adapt-then-Combine (ATC) strategies combined with Least-Mean-Squares (LMS) and Conjugate Gradient (CG) algorithms, outperforming state of the art cooperative localization methods.
    \item More importantly, the proposed diffusion approaches allow each agent to estimate accurately both its position but also the position of the neighboring vehicles facilitating more accurate 4D situational awareness.
    \item The aforementioned methods are evaluated using both well known kinematic models \cite{Thrun2006}, as well as realistic traffic patterns generated by the CARLA autonomous driving simulator \cite{dosovitskiy2017carla}. 
    \item Finally, we investigate the possible impact of VANET size, number of vehicle connections, network delay and range measurements uncertainty to the performance of the various diffusion strategies.
\end{itemize}
Extensive simulation studies verify that: i) Traditional LMS is always outperformed by the other variants, though it provides increased robustness in the presence of range measurements uncertainty ii) LMS and CG with measurements exchanges exhibit similar convergence properties, iii) CG fails to operate adequately both in the presence of network delay and increased range measurements noise, thus making LMS with measurements exchanges the most efficient method.

\subsubsection*{Outline and Notation} The rest of the paper is organized as follows: Section \ref{rel_work} surveys related work to CL and information diffusion on graphs; Section \ref{pf} describes the preliminaries about centralized/distributed optimization and system model; Section \ref{distrloc} presents the proposed distributed and diffusion localization schemes; Section \ref{track} provides a heuristic approach to initialize the diffusion schemes during the tracking mode; Section \ref{sim} is dedicated to experimental setup and evaluation, while Section \ref{conc} concludes our work. The notations used throughout this paper are summarized on TABLE \ref{notation}. 
\begin{table}[htbp]
  \centering
  \caption{Notations}
  \begin{tabular}{r|l}
    \toprule
    $a, \boldsymbol{a}$ and $\boldsymbol{A}$ &  Scalar, vector, and matrix \\
    $\boldsymbol{A}^T$, $\boldsymbol{A}^{-1}$ &  Matrix transpose and matrix inverse\\
    $tr(\boldsymbol{A})$, $\lambda_{max}(\boldsymbol{A})$ & trace and largest eigenvalue of matrix \\
    $t,k$ & time instant and iteration number \\
    $\mathbb{I}_N$ & $N \times N$ identity matrix \\
    $\norm{\cdot}$ & Euclidean norm \\
    $\mathcal{G}\left(\mu, \Sigma \right)$ & Gaussian distribution with mean $\mu$ and covariance $\Sigma$\\
    \bottomrule
  \end{tabular}
  \label{notation}
\end{table}

\section{Related work}
\label{rel_work}
Several works have been proposed in literature that study the benefits and tackle the challenges of CL. An overview of CL in Wireless Sensor Networks is provided in \cite{Buehrer2018}. The CL algorithms can be categorized as: Centralized vs Distributed, One-shot vs Tracking, non-Bayesian vs Bayesian. Usually, distributed approaches are considered to be more attractive, since the computations and processing is performed by each individual vehicle, without any central authority involvement. One-shot and non-Bayesian methods are mostly rely on Maximum Likelihood Estimation (MLE), while Tracking/Bayesian utilize a Minimum Mean Square Error estimator. In \cite{Elazab2017}, a distributed CL method, where each vehicle fuses self and neighbors' absolute  position  from  GPS,  motion  sensor’s readings and range measurements by deploying an Extended Kalman Filter. Round Trip Time is considered as the ranging technique. The approach mainly focuses on tunnels, when GPS may not operating due to signal blockage. In \cite{Soatti2018}, a set of detected non-cooperative features, are used as common noisy reference points. A centralized and distributed scheme based on Bayesian Gaussian Message Passing has been developed, improving stand-alone GPS accuracy in different urban conditions. Both methods exhibited identical performances. A distributed robust cubature Kalman Filter (KF) enhanced  by Huber M-estimation is presented in \cite{Liu2017}. The method is used to tackle the challenges of the data  fusion  under  the  presence of outliers, though considering only a fixed VANET’s size. Pseudo-range measurements from satellites are also considered during the fusion process. A distributed method has been developed in \cite{Rohani2015}, formulating a Bayesian approach which reduces the location estimation uncertainty. A KF is employed, fusing absolute self and neighbors' GPS positions, motion sensors and V2V range measurements. Accuracy of inter-vehicular measurements, size of VANET and communication latency are used to evaluate the effectiveness. 
A distributed non-Bayesian method is presented  in \cite{Kim2018}, which fuses absolute positions, relative distances and angles of only four vehicles, using MLE. The latter is attractive due to its consistency, asymptotic optimality and normality properties.

Information diffusion on graphs approaches facilitate the agents of a network to estimate and learn a common set of parameters in a distributed manner, i.e. location vector of a group of vehicles within a range. Diffusion and distributed estimation algorithms based on LMS have originally been developed in \cite{Cattivelli2010a}. Authors formulated the two standard frameworks of ATC and Combine-then-Adapt (CTA) to perform the distributed processing. As it has been reported, ATC outperforms CTA.
Multi Task Learning (MTL) approaches for estimating a parameter vector not entirely common across the nodes, are highlighted in \cite{Nassif2020}. They are mainly utilize the topology of network graph, in order to create connected sub-graphs and perform the diffusion. Distributed schemes based on Recursive-Least-Sqaures (RLS) and Kalman Filtering have been also proposed in literature \cite{Cattivelli2008}, \cite{Cattivelli2010}. Usually, diffusion approaches require a large number of iterations and message exchanges to converge to the optimal solution. Thus, any communication delay of receiving the necessary information, may have strong impact to the learning performance. To address this issue, which is also evident in any VANET, both LMS and RLS based solutions have been developed \cite{Hua2020}, \cite{Rastegarnia2020}. Although they proved that convergence to the optimal solution can be achieved, this comes at the cost of larger number of iterations. To tackle the limitations of LMS (low convergence speed) and RLS (high complexity and numerical instability) solutions, the authors of \cite{Xu2016} formulated a CG based diffusion scheme, adopting CTA framework. They focus on distributed spectrum estimation and prove the efficiency of their approach.

Distributed CL approaches, as mentioned before, do not evaluate how accurate ego vehicle estimates the location of neighbors, which is a serious limitation towards efficient cooperative awareness. At the same time, information diffusion requires a large number of iterations and message exchange to converge to the optimal solution, which may have a strong impact on time critical applications like cooperative localization.

\section{Preliminaries and System Model}
\label{pf}
\subsection{Centralized Optimization}

\begin{figure}[htbp]
\centerline{\includegraphics[width=0.4\linewidth]{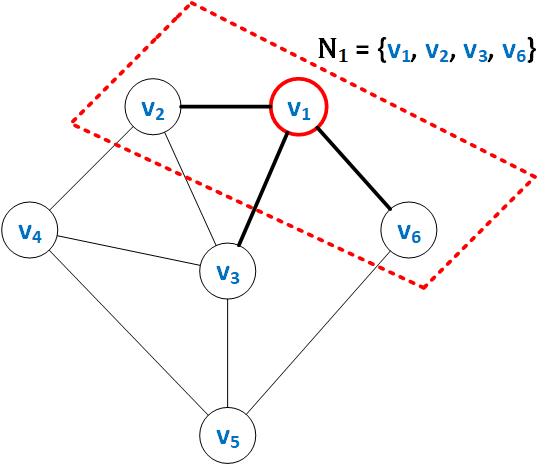}}
\caption{Network of $N$ nodes}
\label{fig1}
\end{figure}

Consider a network graph of $N$ nodes/vehicles (example depicted on Fig.~\ref{fig1}), 
where each node $i$ ($i = 1, \ldots N$) has access to a scalar measurement $d_i^{(t)} \in \mathbb{R}$ and a regression vector $\boldsymbol{u_i^{(t)}} \in \mathbb{R}^N$ at time instant $t$ ($t = 1, \ldots T$). Edges between nodes imply communication connection. 
It is assumed that each node follows a linear measurement model according to:
\begin{align}
\begin{split}
\label{eq:1}
     d_i^{(t)} = \boldsymbol{u_i^{(t)T}w^{(t)}} + n_i^{(t)}
\end{split}
\end{align}

{
As it will be thoroughly explained in the following Sections, the general model of \eqref{eq:1} lends itself for cooperative localization applications, since $d_i^{(t)}, \boldsymbol{u_i^{(t)}}$ and $\boldsymbol{w^{(t)}}$ refer to differential coordinate and V2V neighbors of vehicle $i$, as well as the positions of all VANET's vehicles, respectively. } 
The network's objective is to optimally estimate the, common across the nodes, parameter vector $\boldsymbol{w^{(t)}} \in \mathbb{R}^N$. Note that $n_i^{(t)}$ is a zero mean spatially independent measurement noise term, with variance $\sigma_n^2$. To this end, the following global optimization problem can be defined:
\begin{align}
    \begin{split}
    \label{eq:2}
        \argminl_{\boldsymbol{w^{(t)}}} J(\boldsymbol{w^{(t)}}) = \argminl_{\boldsymbol{w^{(t)}}} \sum_{i=1}^N \left(d_i^{(t)} - \boldsymbol{u_i^{(t)T}w^{(t)}}\right)^2
    \end{split}
\end{align}
It is proven that the optimal vector at time instant $t$ is equal to:
\begin{align}
    \begin{split}
    \label{eq:3}
    \boldsymbol{w^{(t)}} = \left(\sum_{i=1}^N \boldsymbol{R_i^{(t)}}\right)^{-1}\sum_{i=1}^N \boldsymbol{r_i^{(t)}},
    \end{split}
\end{align}
{
with positive semi-definite covariance matrix $\boldsymbol{R_i^{(t)}} = \boldsymbol{u_i^{(t)}u_i^{(t)T}}$ and cross correlation vector $\boldsymbol{r_i^{(t)}} = \boldsymbol{u_i^{(t)T}d_i^{(t)}}$. As it may be obvious, \eqref{eq:3} is actually a centralized (least-squares) implementation since all nodes have to broadcast their measurements and regression vectors to an overall fusion center, which in turn will estimate parameter $\boldsymbol{w^{(t)}}$ and inform nodes about it. However, it is more robust and cost effective for each node to estimate $\boldsymbol{w^{(t)}}$ in a distributed manner and on its own, relying only to its connected neighbors. Distributed implementation of \eqref{eq:2}, in which the proposed Graph Laplacian methods will be derived from, is discussed in the following subsection.
}

\subsection{Distributed Optimization}

A gradient descent based solution will be employed to address the distributed solution of \eqref{eq:2}. For a total number of iterations, say $K$, the optimal vector at each iteration $k$ is given by:
\begin{align}
    \begin{split}
    \label{eq:4}
    \boldsymbol{w_i^{(t,k+1)}} = \boldsymbol{w_i^{(t,k)}} - \mu^{(t)}\frac{\partial{J(\boldsymbol{w^{(t,k)}})}}{\partial {\boldsymbol{w^{(t,k)}}}},
    \end{split}
\end{align}
where $\frac{\partial{J(\boldsymbol{w^{(t,k)}})}}{\partial {\boldsymbol{w^{(t,k)}}}} = \sum_{i=1}^N\left(\boldsymbol{R_i^{(t)}w_i^{(t,k)}} - \boldsymbol{r_i^{(t)}}\right)$  and small scalar step size $\mu^{(t)} > 0$. Thus, \eqref{eq:4} is given by:
\begin{align}
    \begin{split}
    \label{eq:5}
    \boldsymbol{w_i^{(t,k+1)}} = \boldsymbol{w_i^{(t,k)}} - \mu^{(t)}\sum_{i=1}^N\boldsymbol{u_i^{(t)T}}\left(d_i^{(t)} - \boldsymbol{u_i^{(t)}w_i^{(t,k)}}\right)
    \end{split}
\end{align}
Although \eqref{eq:5} is not a distributed implementation since data across the whole network are required, it motivated the development of distributed ATC diffusion LMS algorithm \cite{Cattivelli2010a}:
\begin{align}
    \begin{split}
    \label{eq:6}
    \boldsymbol{\psi_i^{(t,k+1)}} = \boldsymbol{w_i^{(t,k)}} - \mu_i^{(t)}\boldsymbol{u_i^{(t)T}}\left(d_i^{(t)} - \boldsymbol{u_i^{(t)}w_i^{(t,k)}}\right)
    \end{split}\\
    \begin{split}
    \label{eq:7}
    \boldsymbol{w_i^{(t,k+1)}} = \sum_{l \in \mathcal{N}_i^{(t)}}c_{il}^{(t)}\boldsymbol{\psi_l^{(t,k+1)}}
    \end{split}
\end{align}
The neighborhood of node $i$ is the set $\mathcal{N}_i^{(t)}$ with cardinality $\vert{\mathcal{N}_i^{(t)}}\vert$, consisting of self and neighbouring nodes. During the adaptation step of \eqref{eq:6}, each node $i$ estimates in parallel the intermediate vector $\boldsymbol{\psi_i^{(t,k+1)}}$, based on the previously estimated vector $\boldsymbol{w_i^{(t,k)}}$ and the pair $\{d_i^{(t)}, \boldsymbol{u_i^{(t)}}\}$. During the combination step of \eqref{eq:7}, each node $i$ again in parallel, receives the intermediate vectors from its neighbourhood and convexly combines them, in order to estimate the common parameter vector. {That last step is critical for feasible estimation of the parameter vector. Actually without it, the node is unable to estimate accurately the desired vector.} Combination weights $c_{il}^{(t)}$ define the combination matrix $\boldsymbol{C^{(t)}} \in \mathbb{R}^{N \times N}$. A typical choice for combination weights is based on the Metropolis rule \cite{Chouvardas2011}.

\begin{figure}[htbp]
\centerline{\includegraphics[width=0.6\linewidth]{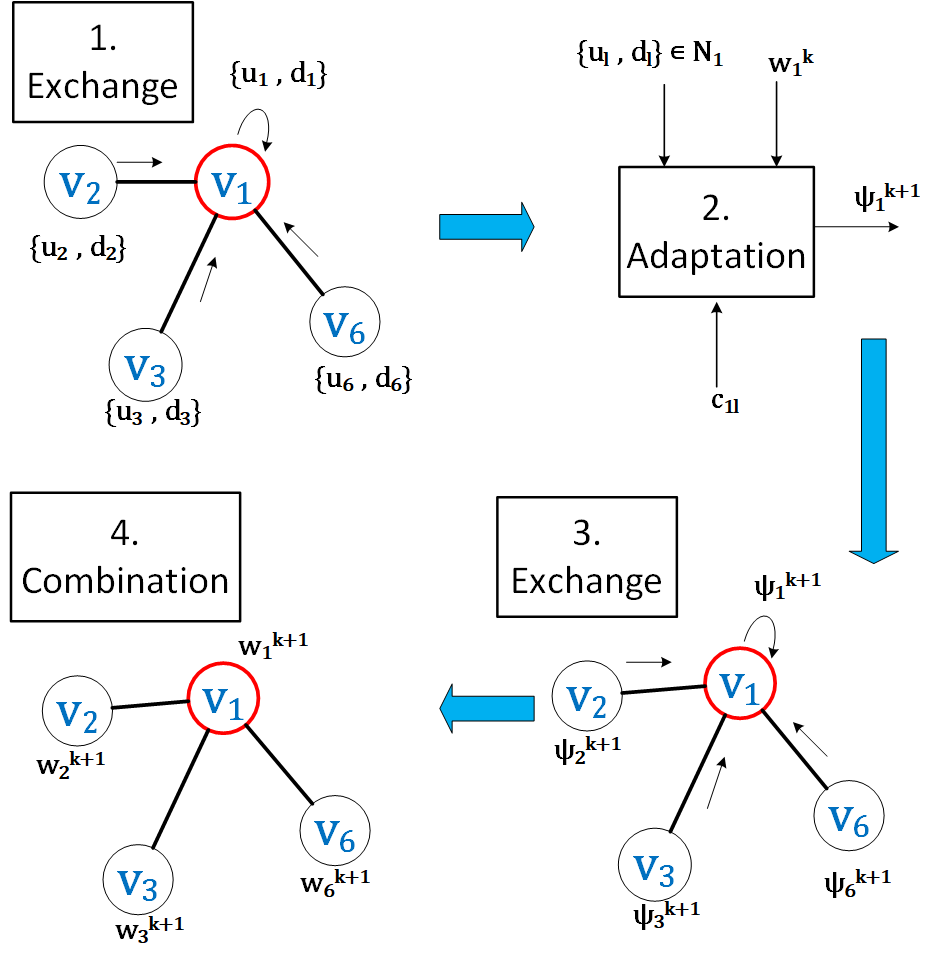}}
\caption{ATC with measurements exchange}
\label{fig2}
\end{figure}

A variant of ATC has also been proposed, where a convex combination operation is added to the adaptation step, leading to ATC with measurements exchanges diffusion LMS:
\begin{align}
    \begin{split}
    \label{eq:8}
    \boldsymbol{\psi_i^{(t,k+1)}} = \boldsymbol{w_i^{(t,k)}} - \mu_i^{(t)}\sum_{l \in \mathcal{N}_i^{(t)}}c_{il}^{(t)}\boldsymbol{u_l^{(t)T}}\left(d_l^{(t)} - \boldsymbol{u_l^{(t)}w_i^{(t,k)}}\right)
    \end{split}\\
    \begin{split}
    \label{eq:9}
    \boldsymbol{w_i^{(t,k+1)}} = \sum_{l \in \mathcal{N}_i^{(t)}}c_{il}^{(t)}\boldsymbol{\psi_l^{(t,k+1)}}
    \end{split}
\end{align}
Notice from \eqref{eq:8} that each node receives now the pair $\{d_l^{(t)}, \boldsymbol{u_l^{(t)}}\}$ from its neighbours, adding one more communication and exchange step to the diffusion algorithm. The LMS diffusion strategy of ATC with measurements exchanges for node v$_1$ is describing on Fig.~\ref{fig2}. Consequently, each node estimates the common parameter vector in a distributed manner, hence avoiding the heavy computational burden of a centralized processing architecture. 

\subsection{System model}
Before we proceed with describing the proposed distributed and diffusion localization strategies, we will shortly review our previous work on \textbf{CLL}. The latter, a graph based approach for CL, performs cooperative multi-modal fusion, exploiting the spatial coherences and connectivity properties of the vehicles of a vehicular network. Furthermore, a measurement model fusing GPS positions, range measurements and the topology of VANET using the linear graph Laplacian operator, is utilized. The main assumption of that approach relies on the feasible estimation of the so-called differential coordinates, by encoding each vehicle's position relative to its neighbouring.

\begin{figure}[htbp]
\centerline{\includegraphics[width=0.7\linewidth]{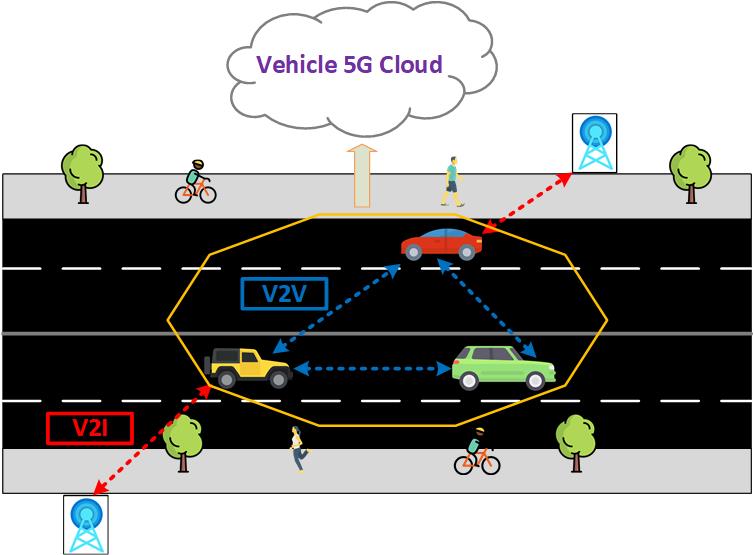}}
\caption{Vehicular ad-hoc network}
\label{fig3}
\end{figure}

Consider a 2-D region where $N$ vehicles of a VANET (shown in Fig.~\ref{fig3}) at an urban environment collect and exchange measurements using V2V communications. 
{Vehicles communicate with the other vehicles of VANET only within a direct communication range $r_c$, according to established V2V communication standards \cite{2018}. That means that vehicles receive messages from other vehicles if only their distance is below than $r_c$. Furthermore, we set a maximum number of closest neighbors $N_{max}$ for each vehicle, as in \cite{Soatti2018}.
Thus, a reduced computational load with permissible localization accuracy can be achieved, avoiding also extreme cases like two vehicles of the same VANET are far (even kilometers) away from each other. Vehicles can utilize a multiple access like communication protocol. As it will be shown in Section \ref{sim}, attained localization accuracy is very promising without the need of all-to-all vehicles connection, with number of vehicles greater than five. Hence, scalability is another important aspect of the proposed approaches.
}

\begin{figure}[htbp]
\centerline{\includegraphics[width=0.6\linewidth]{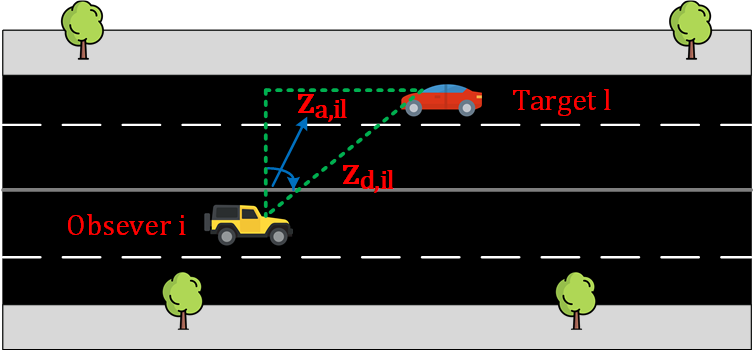}}
\caption{Range measurements}
\label{fig4}
\end{figure}

The absolute location of $i$-th vehicle at time instant $t$ is equal to $\boldsymbol{p_i^{(t)}} = \begin{bmatrix}
x_i^{(t)} &
y_i^{(t)}
\end{bmatrix}^T \in \mathbb{R}^2$, whereas the distance and azimuth angle between connected vehicles $i$ and $l$ are equal to $z_{d,il}^{(t)} = \norm{\boldsymbol{p_i^{(t)}} - \boldsymbol{p_l^{(t)}}}$ and $z_{az,il}^{(t)} = 
\begin{cases}
         \epsilon\pi + \arctan{\frac{\vert x_l^{(t)} - x_i^{(t)}\vert}{\vert y_l^{(t)} - y_i^{(t)}\vert}} , \
         \epsilon = 0,1 \\ 
         \epsilon\pi + \arctan{\frac{\vert y_l^{(t)} - y_i^{(t)}\vert}{\vert x_l^{(t)} - x_i^{(t)}\vert}} , \ \epsilon = \frac{1}{2}, \frac{3}{2}
\end{cases},$ 
shown in Fig.~\ref{fig4} with $\epsilon = 0$.
These measurements could be provided by sensors like GPS, LIDAR, RADAR or Camera, which are integrated to the vehicles, assuming also additive white Gaussian measurement noise \cite{Buehrer2018}. Hence, we acquire the following three measurement models for each vehicle:
\begin{itemize}
    \item Absolute position measurement: 
    \begin{align*}
    \begin{split}
        \boldsymbol{\Tilde{z}_{p,i}^{(t)}} = \boldsymbol{p_i^{(t)}} + \boldsymbol{n_p^{(t)}}, \ \boldsymbol{n_p^{(t)}} \sim \mathcal{G}(0,\boldsymbol{\Sigma{_p}})
    \end{split}
    \end{align*} Covariance matrix $\boldsymbol{\Sigma{_p}}$ is a diagonal matrix equal to $diag(\sigma_x^2, \sigma_y^2)$.
    \item Distance measurement:
    \begin{align*}
    \begin{split}
        \Tilde{z}_{d,il}^{(t)} = z_{d,il}^{(t)} + n_d^{(t)}, \ n_d^{(t)} \sim \mathcal{G}(0,\sigma{_d^2})
    \end{split}
    \end{align*}
    \item Azimuth Angle measurement:
    \begin{align*}
    \begin{split}
       \Tilde{z}_{az,il}^{(t)} = z{_{az,il}^{(t)}} + n_{az}^{(t)}, \ n_{az}^{(t)} \sim \mathcal{G}(0,\sigma{_{az}^2})
    \end{split}
    \end{align*}
\end{itemize}
Moreover, we define the Laplacian matrix of VANET graph $\boldsymbol{L^{(t)}} \in \mathbb{R}^{N \times N}$ as $\boldsymbol{L^{(t)}} = \boldsymbol{D^{(t)}}-\boldsymbol{V^{(t)}}$, where $\boldsymbol{D^{(t)}}, \boldsymbol{V^{(t)}} \in \mathbb{R}^{N \times N}$ the degree and adjacency matrices of VANET graph.
The differential coordinates per vehicle  $\boldsymbol{\delta_i^{(t)}} = \begin{bmatrix}
\delta_i^{(t,x)} & \delta_i^{(t,y)}
\end{bmatrix}^T \in \mathbb{R}^2$ are equal to: 
\begin{align*}
\delta{_i^{(t, x)}} = \frac{1}{\vert{\mathcal{N}_i^{(t)}}\vert-1}{\sum_{l \in \mathcal{N}_i^{(t)}} \left(x_i^{(t)} - x_l^{(t)}\right)} = \\ \frac{1}{\vert{\mathcal{N}_i^{(t)}}\vert-1}{\sum_{l \in \mathcal{N}_i^{(t)}} \left(-\Tilde{z}^{(t)}_{d,il}\sin{\Tilde{z}^{(t)}_{az,il}}\right)},
\end{align*}
and 
\begin{align*}
\delta{_i^{(t, y)}} =  \frac{1}{\vert{\mathcal{N}_i^{(t)}}\vert-1}{\sum_{l \in \mathcal{N}_i^{(t)}} \left(y_i^{(t)} - y_l^{(t)}\right)} = \\ \frac{1}{\vert{\mathcal{N}_i^{(t)}}\vert-1}{\sum_{l \in \mathcal{N}_i^{(t)}} \left(-\Tilde{z}^{(t)}_{d,il}\cos{\Tilde{z}^{(t)}_{az,il}}\right)}
\end{align*}
By creating the vector $\boldsymbol{\delta^{(t,x)}} \in \mathbb{R}^N$ ($x$-differentials), the $x$-coordinates position of vehicles (as vector $\boldsymbol{x^{(t)}} \in \mathbb{R}^N$) follow the linear measurement model:
\begin{align}
    \begin{split}
    \label{eq:10}
        \boldsymbol{L^{(t)}}\boldsymbol{x^{(t)}} = \boldsymbol{D^{(t)}}\boldsymbol{\delta{^{(t, x)}}},
    \end{split}
\end{align}
However, $\boldsymbol{L^{(t)}}$ is positive semi-definite and non-invertible \cite{Sorkine2005}, and as such \eqref{eq:10} is reformulated by creating, the extended Laplacian matrix $\boldsymbol{\Tilde{L}^{(t)}} \in \mathbb{R}^{2N \times N}$ as $\boldsymbol{\Tilde{L}^{(t)}} = \begin{bmatrix}
\boldsymbol{L^{(t)}} \\ \hline
\mathbb{I}_{N}
\end{bmatrix}$ and vector $\boldsymbol{s^{(t,x)}} \in \mathbb{R}^{2N}$ as $\boldsymbol{s^{(t,x)}} = \begin{bmatrix}
\boldsymbol{D^{(t)}}\boldsymbol{\delta^{(t,x)}} \\ \hline
\boldsymbol{\Tilde{z}_p^{(t,x)}}
\end{bmatrix}$, where $\boldsymbol{\Tilde{z}_p^{(t,x)}} \in \mathbb{R}^N$ are the GPS $x$-positions of vehicles, which act as anchor points \cite{Sorkine2005}. Therefore, \eqref{eq:10} transforms to:
\begin{align}
    \begin{split}
    \label{eq:11}
        \boldsymbol{\Tilde{L}^{(t)}}\boldsymbol{x^{(t)}} = 
         \boldsymbol{s{^{(t, x)}}},
    \end{split}
\end{align} 
which can be solved in the least-squares sense and estimate the $x$-positions of vehicles. Obviously, the same approach is followed for $y$-positions $\boldsymbol{y^{(t)}} \in \mathbb{R}^N$. Note that we make an explicit statement about data association, i.e. for every range measurement ($\tilde{z}_{d,il}^{(t)}$ or $\tilde{z}_{az,il}^{(t)}$) of vehicle $i$, the latter knows which vehicle $l$ is associated with. Although it is possible to model the associations by comparing range measurements with GPS positions of neighbors, we assume that data association is given to us as part of a pre-processing step.
Thus, \textbf{CLL} assumes that communication and sensing graph are identical.

However, \eqref{eq:11} is a centralized implementation, since the vehicles are required to broadcast and send their ids and measurements to a central node/fusion center (e.g. 5G cloud), which in turn will estimate and inform them about their positions. This is exactly what should be avoided due to serious limitations and drawbacks of central node operation like malfunctioning or cyber-attack. \textbf{CLL} will act as a baseline to the development of fully distributed solutions for CL. 

\section{Distributed and Diffusion Localization strategies}
\label{distrloc}
In this Section, the proposed distributed and diffusion localization methods will be derived based on LMS and CG algorithms. All three methods exploit the ATC framework due to its superiority over CTA, and aim not only to converge to the global \textbf{CLL} solution, but to outperform it by means of information diffusion, utilizing the graph Laplacian operator. 
\subsection{Graph Laplacian LMS}
It is important to notice the resemblance between \eqref{eq:10} and \eqref{eq:1}. In both cases, the vector to be estimated is linearly correlated with measurements and regression vectors. In the localization framework, the differential coordinates, divided by nodes' degrees, act as the measurements, while the rows of Laplacian matrix as the corresponding regression vectors. Subsequently, for each vehicle, the following linear measurement model (for $x$-coordinates) is defined:
\begin{align}
    \begin{split}
    \label{eq:12}
        \delta_i^{(t,x)} = \boldsymbol{L^{(t)T}_{i:}}\boldsymbol{w_{i,x}^{(t)}},
    \end{split}
\end{align}
where the unknown vector $\boldsymbol{w_{i,x}^{(t)}} \in \mathbb{R}^N$ corresponds to the $x$-coordinates position of vehicles, estimated by vehicle $i$. Apparently, the optimal vector is common across the vehicles of the network. Based on that local model, each vehicle is totally unable to estimate its position and all others on its own. Only by means of cooperation and information diffusion through V2V and 5G related protocols, the vehicles can learn the desired vector. As a natural consequence of \eqref{eq:6} and \eqref{eq:7}, the two steps of the proposed \textbf{Graph Laplacian LMS} or \textbf{GLLMS} approach are derived:
\begin{align}
    \begin{split}
    \label{eq:13}
    \boldsymbol{\psi_{i,x}^{(t,k+1)}} = \boldsymbol{w_{i,x}^{(t,k)}} - \mu_{1,i}^{(t)}\boldsymbol{L_{i:}^{(t)T}}\left(\delta_i^{(t,x)} - \boldsymbol{L_{i:}^{(t)}w_{i,x}^{(t,k)}}\right)
    \end{split}
\end{align}
\begin{align}
    \begin{split}
    \label{eq:14}
    \boldsymbol{w_{i,x}^{(t,k+1)}} = \sum_{l \in \mathcal{N}_i^{(t)}}c_{il}^{(t)}\boldsymbol{\psi_{l,x}^{(t,k+1)}}
    \end{split}
\end{align}
At the adaptation step of \eqref{eq:13}, each vehicle estimates in parallel the intermediate vectors, using the pair $\{\delta_i^{(t,x)}, \boldsymbol{L_{i:}^{(t)T}} \}$, while in the combination step of \eqref{eq:14}, receives from and sends to the neighborhood the intermediate vectors, in order to estimate the desired vectors. At the end of that procedure, each vehicle will estimate and converge to the same location vector. 
During the initialization stage at time instant $t$, the desired vector is set to the GPS positions of the vehicles, as a rough estimation of the solution vector. Vehicles via, e.g. TDMA protocol, are in fact informed about the noisy locations of all VANET's vehicles (not just neighbors), in order to successively estimate their positions.
The \textbf{GLLMS} approach for $x$-coordinates is summarized on \textbf{Algorithm \ref{GLLMS}}.

It is worth mentioning that each vehicle is interested enough to not only estimate accurately its own position, but also its direct neighborhood and all others vehicles of the network. Since individual vehicles would estimate on its own the entire location vector, the distributed and diffusion localization will facilitate the design of an efficient individual path planning and control mechanism. The latter will determine the best possible future driving actions, improving the overall performance of VANET in terms of e.g. reduced traffic accidents and fuel consumption. 
\begin{algorithm}
\KwIn{$N$, $T$, $K$, $\boldsymbol{\Tilde{z}_{p}^{(t,x)}}$, $\delta_i^{(t,x)}$, $\boldsymbol{L_{i:}^{(t)}}$, $\boldsymbol{C^{(t)}}$}
\KwOut{$\boldsymbol{w_{i,x}^{(t,K)}} \in \mathbb{R}^N,  \forall i \in N$}
\For{$t=1,\ldots T$}{
  \For {each vehicle $i$ in parallel}{
   $\boldsymbol{w_{i,x}^{(t,1)}}$ = $\boldsymbol{\Tilde{z}_{p}^{(t,x)}}$\;
   \For {$k = 1,\ldots K$}{
   $\mu_{1,i}^{(t)}$ from \eqref{m1}\;
   $\boldsymbol{\psi_{i,x}^{(t,k+1)}} = \boldsymbol{w_{i,x}^{(t,k)}} - \mu_{1,i}^{(t)}\boldsymbol{L_{i:}^{(t)T}}\left(\delta_i^{(t,x)} - \boldsymbol{L_{i:}^{(t)}w_{i,x}^{(t,k)}}\right)$\; 
   $\boldsymbol{w_{i,x}^{(t,k+1)}} = \sum_{l \in \mathcal{N}_i^{(t)}}c_{il}^{(t)}\boldsymbol{\psi_{l,x}^{(t,k+1)}}$\;
   }
}
}
\caption{\textbf{Graph Laplacian LMS or GLLMS}} \label{GLLMS}
\end{algorithm}
\subsection{Graph Laplacian CG for improved cooperative awareness}
The previous distributed and diffusion localization method converges each time instant to the global solution at the cost of high enough total number of iterations. This is a serious limitation towards real-time implementation, since the vehicles are required to estimate their positions before the new GPS measurement arrive, namely between 100-300 ms. Since vehicles are connected through V2V, it is expected that an additional measurement exchange step, will speed up the convergence of diffusion. Usually, CG algorithm is employed to accelerate LMS type approaches. The acceleration is attained due to the fact that CG selects the successive direction vectors towards solution as a conjugate version of the
successive gradients obtained at each step. The conjugate directions are not specified beforehand, but rather are
determined sequentially at each step of the iteration. The directions are based on the gradients, therefore the process makes
good uniform progress toward the solution at every step. Furthermore, no line searching is required at any stage as in LMS. 

We define matrix $\boldsymbol{U_i^{(t)}} \in \mathbb{R}^{\vert \mathcal{N}_i^{(t)} \vert \times N}$ and vector $\boldsymbol{q_i} \in \mathbb{R}^{\vert \mathcal{N}_i^{(t)} \vert}$ by putting together the rows of Laplacian matrix and differential coordinates, respectively, which belong to the neighborhood $\mathcal{N}_i$. As a matter of fact, neighboring vehicles broadcast the pair $\{\boldsymbol{L_{l:}^{(t)T}}, \delta_l^{(t,x)} \}$, vehicle $i$ receives them and together with its own data, defines $\{\boldsymbol{U_i^{(t)}}, \boldsymbol{q_i^{(t)}}\}$. The first row/element of the pair corresponds to $i$-th vehicle. Thus, the following linear problem can be defined:
\begin{align}
    \begin{split}
    \label{eq:LMS_prob}
    \boldsymbol{U_i^{(t)}}\boldsymbol{w_{i,x}^{(t)}} = \boldsymbol{q_i^{(t)}}
    \end{split}
\end{align}
Furthermore, if we utilize the instantaneous positive semi-definite covariance matrix $\boldsymbol{A_i^{(t)}} = \boldsymbol{U_i^{(t)T}}\boldsymbol{U_i^{(t)}}$, $\boldsymbol{A_i^{(t)}} \in \mathbb{R}^{N\times N}$ and covariance vector $\boldsymbol{b_i^{(t)}} = \boldsymbol{U_i^{(t)T}}\boldsymbol{q_i^{(t)}}$, $\boldsymbol{b_i^{(t)}} \in \mathbb{R}^{N}$, we end up (as a direct sequence of \eqref{eq:LMS_prob}) to the following linear problem:
\begin{align}
    \begin{split}
    \label{eq:17}
    \boldsymbol{A_i^{(t)}}\boldsymbol{w_{i,x}^{(t)}} = \boldsymbol{b_i^{(t)}}
    \end{split}
\end{align}
The CG optimization method is used to tackle \eqref{eq:17}, avoiding the high complexity and unstable performance of RLS-type methods. Its main idea lies on the fact that a set of direction vectors conjugate to $\boldsymbol{A_i^{(t)}}$ are exploited to estimate the desired optimal vector $\boldsymbol{w_{i,x}^{(t)}}$. 

However, \eqref{eq:17} corresponds also to the minimization of cost function \cite{Lalos2008} $\mathcal{V}\left(\boldsymbol{w_{i,x}^{(t)}}\right) = \mathbb{E}\{\boldsymbol{(q_i^{(t)})^2}\} - \boldsymbol{b_i^{(t)T}}\boldsymbol{w_{i,x}^{(t)}} - \boldsymbol{w_{i,x}^{(t)T}}\boldsymbol{b_i^{(t)}} + \boldsymbol{w_{i,x}^{(t)T}}\boldsymbol{A_i^{(t)}}\boldsymbol{w_{i,x}^{(t)}}$. The main steps of the proposed
\textbf{Graph Laplacian CG} or \textbf{GLCG} are summarized on \textbf{Algorithm \ref{GLCG}}.
\begin{algorithm}
\KwIn{$N$, $T$, $K$, $\boldsymbol{\Tilde{z}_{p}^{(t,x)}}$, $\boldsymbol{U_i^{(t)}}$, $\boldsymbol{q_i^{(t)}}$ $\boldsymbol{C^{(t)}}$, $\lambda$}
\KwOut{$\boldsymbol{w_{i,x}^{(t,K)}} \in \mathbb{R}^N,  \forall i \in N$}
\For{$t=1,\ldots T$}{
  \For {each vehicle $i$ in parallel}{
   $\boldsymbol{w_{i,x}^{(t,1)}}$ = $\boldsymbol{\Tilde{z}_{p}^{(t,x)}}$\;
   $\boldsymbol{A_i^{(t)}} = \boldsymbol{U_i^{(t)T}}\boldsymbol{U_i^{(t)}} + 10^{-7} \cdot \mathbb{I}_N$\; $\boldsymbol{b_i^{(t)}}$ = $\boldsymbol{U_i^{(t)T}}\boldsymbol{q_i^{(t)}}$\;
   $\boldsymbol{g_i^{(1)}} = \boldsymbol{b_i^{(t)}} - \boldsymbol{A_i^{(t)}}\boldsymbol{w_{i,x}^{(t,1)}}$\;
   $\boldsymbol{r_i^{(2)}}$ = $\boldsymbol{g_i^{(1)}}$\;
   \For {$k = 1,\ldots K$}{
   $\alpha_i = \frac{\boldsymbol{r_i^{(k+1)T}}\boldsymbol{g_i^{(k)}}}{\boldsymbol{r_i^{(k+1)T}}\boldsymbol{A_i^{(t)}}\boldsymbol{r_i^{(k+1)}}}$\;
   $\boldsymbol{\psi_{i,x}^{(t,k+1)}} = \boldsymbol{w_{i,x}^{(t,k)}} + \alpha_i\boldsymbol{r_i^{(k+1)}}$\; 
   $\boldsymbol{\tilde{g}_i^{(k)}} = \lambda\boldsymbol{g_i^{(k)}} + \boldsymbol{b_i^{(t)}}- \boldsymbol{A_i^{(t)}}\boldsymbol{w_{i,x}^{(t,k)}}$\;
   $\boldsymbol{g_i^{(k+1)}} = \boldsymbol{\tilde{g}_i^{(k)}} + \alpha_i\boldsymbol{r_i^{(k+1)}}$\;
   $\beta_i = min(\frac{(\boldsymbol{g_i^{(k+1)}}-\boldsymbol{g_i^{(k)}})^T\boldsymbol{g_i^{(k+1)}}}{\boldsymbol{g_i^{(k+1)T}}\boldsymbol{g_i^{(k+1)}}}, \frac{\boldsymbol{g_i^{(k+1)T}}\boldsymbol{g_i^{(k+1)}}}{\boldsymbol{g_i^{(k+1)T}}\boldsymbol{g_i^{(k+1)}}})$\;
   $\boldsymbol{r_i^{(k+2)}} = \boldsymbol{g_i^{(k+1)}} + \beta_i\boldsymbol{r_i^{(k+1)}}$\;
   $\boldsymbol{w_{i,x}^{(t,k+1)}} = \sum_{l \in \mathcal{N}_i^{(t)}}c_{il}^{(t)}\boldsymbol{\psi_{l,x}^{(t,k+1)}}$\;
   }
}
}
\caption{\textbf{Graph Laplacian CG or GLCG}}
\label{GLCG}
\end{algorithm}
Optimal step sizes $\alpha_i$ are computed as the minimizing arguments of $\mathcal{V}$, factors $\beta_i$ are to be ensure $\boldsymbol{A_i^{(t)}}$-orthogonality for the direction vectors $\boldsymbol{r_i}$, while $\boldsymbol{g_i}$ is the negative gradient of $\mathcal{V}$. Factors $\beta_i$ are chosen as the minimum between the Polak-Ribiere formula $\beta_i^{PR} = \frac{(\boldsymbol{g_i^{(k+1)}}-\boldsymbol{g_i^{(k)}})^T\boldsymbol{g_i^{(k+1)}}}{\boldsymbol{g_i^{(k+1)T}}\boldsymbol{g_i^{(k+1)}}}$ and Fletcher-Reeves $\beta_i^{FR} = \frac{\boldsymbol{g_i^{(k+1)T}}\boldsymbol{g_i^{(k+1)}}}{\boldsymbol{g_i^{(k+1)T}}\boldsymbol{g_i^{(k+1)}}}$, in order to avoid anomalous behaviour and numerical instability. A forgetting factor $0 < \lambda < 1$ is employed to update the instantaneous covariance vector. We choose $\lambda = 0.2$. Note that  small factor $10^{-7} \cdot \mathbb{I}_N$ has been added to  $\boldsymbol{A_i^{(t)}}$, since the latter is in fact an ill-conditioned and low-rank matrix, as a product of rows of singular Laplacian matrix. This a serious limitation of the optimization method, since the convergence speed is determined by the condition number $\kappa({\boldsymbol{A_i^{(t)}}}) = $ $\kappa({\boldsymbol{U_i^{(t)T}}\boldsymbol{U_i^{(t)}}}) = $ $\kappa({\boldsymbol{U_i^{(t)}}})^2$: the larger $\kappa({\boldsymbol{A_i^{(t)}}})$ is, the slower the improvement towards solution \cite{Saad2003}. However, by means of information diffusion, \textbf{GLCG} finally succeeds to converge to the optimal solution vector as it is shown in Section 
\ref{sim}. Note that if $\boldsymbol{U_i^{(t)}}$ were chosen to be $\boldsymbol{L_{i:}^{(t)}}$, i.e. avoiding to use measurements exchanges, then $\boldsymbol{A_i^{(t)}}$ would be rank-one matrix, almost prohibitive to be used in estimating the location vector.
 
Therefore, a novel distributed localization scheme based on ATC framework and CG optimization with measurements exchanges has been developed.
Each vehicle creates the matrix/vector $\{\boldsymbol{U_i^{(t)}}, \boldsymbol{q_i^{(t)}}\}$, by receiving the transmitted $\{\boldsymbol{L_{l:}^{(t)T}}, \delta_l^{(t,x)}\}$ from its neighborhood. Afterwards, it estimates the intermediate vectors using CG method and exploits its benefits with contrast to RLS. Finally, it estimates the desired location vector by a convex combination of neighboring intermediate vectors.

Although the communication burden is now increased, it can be performed in efficient manner, since each vehicle has to broadcast only a scalar value and a sparse vector with non-zero integer entries equal to $\vert\mathcal{N}_i^{(t)}\vert$. Laplacian matrix is actually sparse, since vehicles are connected only to a subset of operating vehicles. As it will be shown in Section \ref{sim}, the measurements exchanges step has a significant impact on the convergence speed of the proposed schemes.

\subsection{Graph Laplacian LMS for improved cooperative awareness}
The main limitation of \textbf{GLCG} is related to the ill-conditioned $\boldsymbol{A_i^{(t)}}$, which may deteriorate the performance of diffusion, as well as optimal step sizes $\alpha_i$ and $\beta_i$. Since the latter are estimated directly from available data, they are also vulnerable to increased noisy data, coming either from uncertain measurements or network latency (past estimations treated as current). As such, a variant of \textbf{GLCG} countering those drawbacks, which in addition will act comparatively, should be developed. Diffusion LMS solution of \eqref{eq:LMS_prob} derives exactly  from \eqref{eq:8} and \eqref{eq:9}, formulating the proposed method of \textbf{Graph Laplacian LMS with measurements exchanges} or \textbf{GLLME}:
\begin{align*}
    \begin{split}
    \boldsymbol{\psi_{i,x}^{(t,k+1)}} =
    \end{split}
\end{align*}
\begin{align}
    \begin{split}
    \label{eq:15}
    \boldsymbol{w_{i,x}^{(t,k)}} - \mu_{2,i}^{(t)}\sum_{l \in \mathcal{N}_i^{(t)}}c_{il}^{(t)}\boldsymbol{L_{l:}^{(t)T}}\left(\delta_l^{(t,x)} - \boldsymbol{L_{l:}^{(t)}w_{i,x}^{(t,k)}}\right)
    \end{split}\\
    \begin{split}
    \label{eq:16}
    \boldsymbol{w_{i,x}^{(t,k+1)}} = \sum_{l \in \mathcal{N}_i^{(t)}}c_{il}^{(t)}\boldsymbol{\psi_{l,x}^{(t,k+1)}}
    \end{split}
\end{align}
During the adaptation step of \eqref{eq:15}, an extra communication step has been added, since each vehicle receives from connected neighbors the pair of $l$-th row of Laplacian matrix and the $l$-th differential coordinate, in the form of $\{\delta_l^{(t,x)}, \boldsymbol{L_{l:}^{(t)T}} \}$. 
At the initialization stage, once again the noisy GPS locations are utilized as a rough estimation. The \textbf{GLLME} approach is summarized on \textbf{Algorithm \ref{GLLME}}.

Therefore, both \textbf{GLCG} and \textbf{GLLME} are actually addressing the same optimization problem of \eqref{eq:LMS_prob}, employing CG and LMS with measurements exchanges algorithms, respectively. They mainly focus on improving cooperative awareness ability via the integration of additional information, with respect to \textbf{GLLMS}. However, \textbf{GLLME} doesn't utilize the ill-conditioned $\boldsymbol{A_i^{(t)}}$. At the same time, the optimal step size of the proposed scheme, as well as that of \textbf{GLLMS}, can be optimally determined according to the best practices, exploiting the connectivity topology of vehicles (shown in the next subsection). Thus, avoiding the impact of highly contaminated by noise heterogeneous data to the performance of Laplacian diffusion. Finally, note that the communication overhead of \textbf{GLLME} is the same as \textbf{GLCG}, since the same data pair has to be broadcast.
\begin{algorithm}
\KwIn{$N$, $T$, $K$, $\boldsymbol{\Tilde{z}_{p}^{(t,x)}}$, $\delta_i^{(t,x)}$, $\boldsymbol{L_{i:}^{(t)}}$, $\boldsymbol{C^{(t)}}$}
\KwOut{$\boldsymbol{w_{i,x}^{(t,K)}} \in \mathbb{R}^N,  \forall i \in N$}
\For{$t=1,\ldots T$}{
  \For {each vehicle $i$ in parallel}{
   $\boldsymbol{w_{i,x}^{(t,1)}}$ = $\boldsymbol{\Tilde{z}_{p}^{(t,x)}}$\;
   $\mu_{2,i}^{(t)}$ from \eqref{m2}\;
   \For {$k = 1,\ldots K$}{
   $\boldsymbol{\psi_{i,x}^{(t,k+1)}} = \boldsymbol{w_{i,x}^{(t,k)}} - \mu_{2,i}^{(t)}\sum_{l \in \mathcal{N}_i^{(t)}}c_{il}^{(t)}\boldsymbol{L_{l:}^{(t)T}}\left(\delta_l^{(t,x)} - \boldsymbol{L_{l:}^{(t)}w_{i,x}^{(t,k)}}\right)$\; 
   $\boldsymbol{w_{i,x}^{(t,k+1)}} = \sum_{l \in \mathcal{N}_i^{(t)}}c_{il}^{(t)}\boldsymbol{\psi_{l,x}^{(t,k+1)}}$\;
   }
}
}
\caption{\textbf{Graph Laplacian LMS with measurements exchanges or GLLME}} 
\label{GLLME}
\end{algorithm}

\subsection{Optimal step size of Graph Laplacian LMS}

The regression vector $\boldsymbol{L_{i:}^{(t)}}$ is a deterministic quantity, since is referring to $i$-th row of Laplacian matrix. This property facilitates the optimal selection of step sizes $\mu_{1,i}^{t)}$ and $\mu_{2,i}^{(t)}$, respectively, for convergence in the mean sense \cite{Sayed2008}:
$0 < \mu_{1,i}^{(t)} < \frac{2}{\lambda_1^{max}}$ and 
$0 < \mu_{2,i}^{(t)} < \frac{2}{\lambda_2^{max}}$.
The $\lambda_1^{max}$ corresponds to the maximum eigenvalue of instantaneous covariance matrix $\boldsymbol{L_{i:}^{(t)T}}\boldsymbol{L_{i:}^{(t)}}$, while $\lambda_2^{max}$ corresponds to \cite{Cattivelli2010a} the maximum eigenvalue of instantaneous covariance positive semi-definite matrix  $\sum_{l \in \mathcal{N}_i^{(t)}}c_{il}^{(t)}\left(\boldsymbol{L_{l:}^{(t)T}}\boldsymbol{L_{l:}^{(t)}}\right)$. At the same time, maximum convergence speed is attained when: $\mu_{1,i}^{(t)} = \frac{2}{\lambda_1^{max} + \lambda_1^{min}}$ and $\mu_{2,i}^{(t)} = \frac{2}{\lambda_2^{max} + \lambda_2^{min}}$, where $\lambda_1^{min}, \lambda_2^{min}$ are the minimum eigenvalues of corresponding covariance matrices. 
\begin{proposition} To ensure fast convergence to the optimal solution and asymptotic unbiasedness of $\textbf{GLLMS}$ and $\textbf{GLLME}$, the following must hold:
\begin{align*}
    \begin{split}
    \mu_{1,i}^{(t)} = 
    2/ (\vert\mathcal{N}_i^{(t)}\vert-1)^2 + \vert\mathcal{N}_i^{(t)}\vert-1)
    \end{split}\\
    \begin{split}
    \mu_{2,i}^{(t)} = 
    2/ \lambda_2^{max}
    \end{split}
\end{align*}
\begin{proof}
See Appendix.
\end{proof}
\end{proposition}
Furthermore, we set: 
\begin{align}
    \begin{split}
    \label{m1}
    \mu_{1,i}^{(t)} = min(0.1, \ 2/( (\vert\mathcal{N}_i^{(t)}\vert-1)^2 + \vert\mathcal{N}_i^{(t)}\vert-1))
    \end{split}\\
    \begin{split}
    \label{m2}
    \mu_{2,i}^{(t)} = min\left(0.1, \ 
     2/ \lambda_2^{max}\right),
    \end{split}
\end{align}
in order to avoid large step sizes which may slower down the convergence.
Finally, the convergence in the mean sense of \textbf{GLLME} is guaranteed as follows:
\begin{proposition} A sufficient condition of asymptotic unbiasedness of \textbf{GLLME} is provided by:
\begin{align*}
    0 < \mu_{2,i}^{(t)} < \frac{2}{\argmaxl_{l \in \mathcal{N}_i^{(t)}}( (\vert\mathcal{N}_l^{(t)}\vert-1)^2 + \vert\mathcal{N}_l^{(t)}\vert-1)}
\end{align*}
\end{proposition}
\begin{proof}
See Appendix.
\end{proof}
{
\subsection{Discussion}

Graph Laplacian localization methods address the case of highly dynamic topologies, though only inside the specific VANET. Each vehicle interacts with the members of its own neighborhood $\mathcal{N}_i^{(t)}$, which doesn't necessarily remain always the same. For example, neighboring vehicles (and V2V connections) to $i$ vary while they are moving. However, when some vehicles enter or exit the VANET, Graph Laplacian diffusion schemes should be reset and reinitialized using e.g. GPS, since up to that moment vehicles were estimating the positions of vehicles of the "old" VANET. A mechanism which is adaptable to those inter-VANETs modifications, without the need of locations reset, is of future investigation. 

Another important issue is about data association (we briefly commented on that in Section \ref{pf}). Vehicles use LIDAR or Cameras to extract their distance and angle with respect to other vehicles. However, a data association step is required in order to match range measurements with the correct vehicle id (known from V2V communication). Under the assumption that vehicles detect (through LIDAR) only those vehicle which they do have V2V connection, Graph Laplacian localization features a competitive advantage: no explicit data association is necessary, since the differential coordinates (measurement vector) correspond actually to the average of range measurements.
}

\section {Facilitating Tracking via Efficient Initialization Strategies }
\label{track}
{
Kalman filter is one of the most prominent tracking methods for realizing multi-modal fusion. It usually treats the different measurement models independently in order to estimate the locations. On the contrary, the proposed Graph Laplacian methods facilitate the compact and unified fusion of pair-wise measurements in a single and linear way, via the Laplacian operator, offering some very promising results as shown in Section \ref{sim}. In the context of distributed and diffusion localization, the proposed Graph Laplacian methods could be potentially part of the measurement model of diffusion Kalman. However, major limitations of diffusion Kalman filter are related to the additional and expensive exchange of matrices through the established V2X communication links \cite{Cattivelli2010}. Therefore, the  exploitation of either a Kalman approach or even the prediction step of a Kalman filter by utilizing a realistic kinematic model may be useful, while the proposed scheme can be deployed as an alternative to the estimation step of Kalman. The experiments that we provide highlight the benefits as compared to a traditional least-squares estimation that is usually deployed. Therefore it is expected that the proposed approach can improve the efficiency of a traditional Kalman solution. }

An alternative option for the tracking mode of the proposed diffusion schemes would be their initialization at each time instant to the estimated location vector of the previous time instant, instead of GPS measurements. Given the fact that consecutive locations are quite close to each other, tracking mode is expected to improve the convergence speed. However, between consecutive time instances (i.e. 100-300 ms), distance and angle (in the form of differential coordinates) measurements rarely modify rapidly. Moreover, it is not highly probable the topology of the VANET, i.e. which vehicles are V2V connected, to be changed. As a matter of fact, we claim that using more or less the same measurement and regression pair for a number of time instances, the convergence to the desired solution can't be achieved, rather than remain to the already estimated location vector. To overcome that limitation, we utilize a number of anchor points, i.e. noisy GPS positions, in order to deliberately degrade the solution vector used in the initialization stage. These anchor points provide the diffusion schemes the ability to converge, just like they achieve the solution to the global problem of \eqref{eq:11}. To this end, we have developed a heuristic approach to initialize the proposed diffusion localization schemes each time instant. 
Self-position, as well as non-neighbors, are initialized to their current noisy GPS measurements. The previously estimated location vector initializes half of the neighbors. GPS measurements are also used for the other half of connected vehicles. It is assumed that each vehicle knows the GPS locations of the rest vehicles of VANET. The heuristic initialization approach is summarized on \textbf{Algorithm \ref{Init}}. Lines 3 and 5 account for the self-vehicle and its non-neighbors case. Line 9 addresses the case in which vehicle $i$ is connected to a relatively small (with respect to VANET's size) number of vehicles. Clearly, a suitable choice of $threshold$ depends upon the number of connections, as well as how erroneous the GPS measurement actually is. For example, a more accurate GPS position implies a lower $threshold$.  We choose $threshold = 0.8$. In Lines 12-13 it is chosen that the first half of neighbors of $i$ are initialized to the previous solution, while the other half to their GPS measurements, in order to achieve greater performance in terms of location accuracy.
\begin{algorithm}
\KwIn{$N$, $T$, $K$ $\boldsymbol{\Tilde{z}_{p}^{(t,x)}}$}
\KwOut{$\boldsymbol{w_{i,x}^{(t,1)}} \in \mathbb{R}^N,  \forall i \in N$}
\For{$t=1,\ldots T$}{
  \For {each vehicle $i$ in parallel}{
   $w_{ii,x}^{(t,1)}$ = $\Tilde{z}_{p,i}^{(t,x)}$\;
   \For {each vehicle $j$ not connected to $i$}{
   $w_{ij,x}^{(t,1)}$ = $\Tilde{z}_{p,j}^{(t,x)}$\;
    }
    \eIf{$\frac{|\mathcal{N}_i^{(t)}|}{N} < threshold$}{
    \For {each vehicle $l$ connected to $i$}{
    $w_{il,x}^{(t,1)}$ = $w_{il,x}^{(t-1,K)}$\;}}{
    Initialize first half of neighbors to previous solution $w_{il,x}^{(t-1,K)}$\;
    Initialize the other half of neighbors to GPS measurement $\Tilde{z}_{p,l}^{(t,x)}$\;
    
    }
    
   }
}
\caption{\textbf{Coherent initialization}}
\label{Init}
\end{algorithm}

\section{Simulations}
\label{sim}
In this Section, we validate the introduced approaches by performing computer based simulations using Python and CARLA simulator. { During the experiments a PC Laptop with 8GB RAM and Intel Core i7-1065G7 CPU @ 1.3 GHz was used.}

\subsection{Experimental Setup}
\begin{figure}[htbp]
  \centering
  \subfloat[Ground truth]{\includegraphics[width=0.45\linewidth]{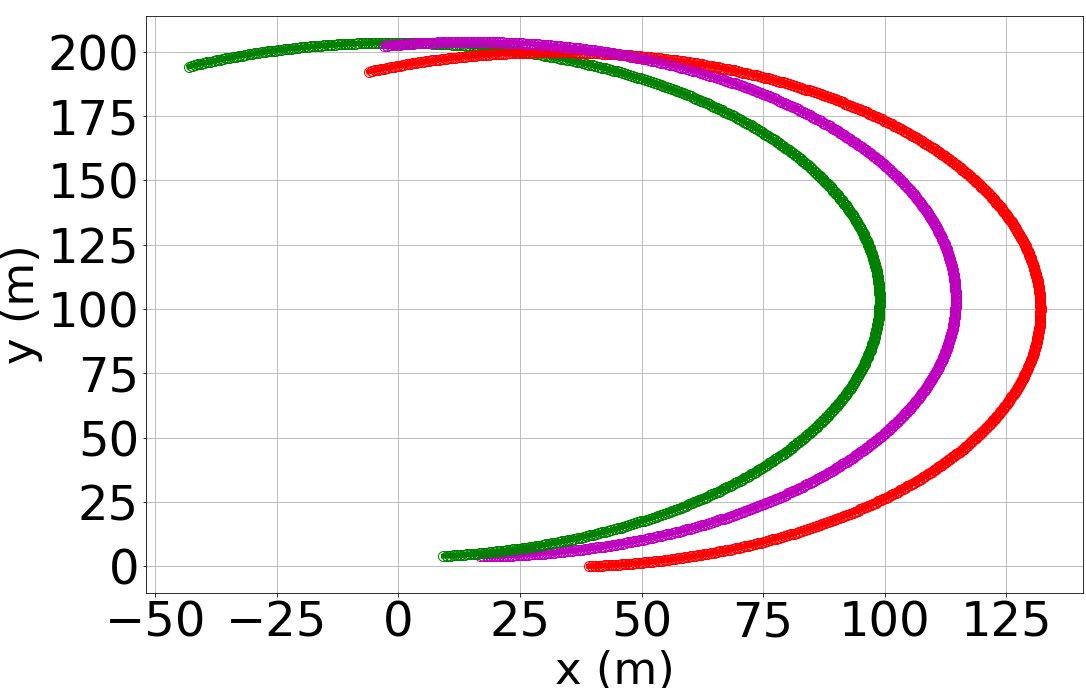}}
  \subfloat[VANET graph example]{\includegraphics[width=0.45\linewidth]{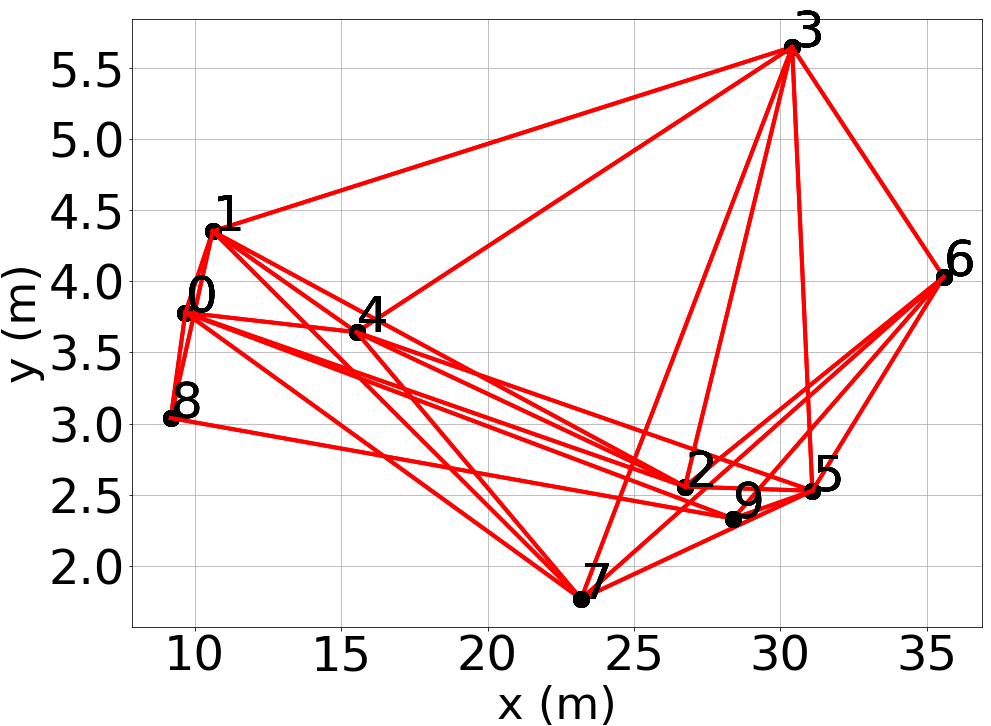}}
  \caption{Reference trajectories and VANET graph}
  \label{traj-graph}
\end{figure}

CARLA simulator has been employed to extract different traffic patterns of vehicles moving in an urban city (example shown in Fig.~\ref{traj-CARLA}).
We generated also the positions of $N$ vehicles using the bicycle kinematic model \cite{Thrun2006}. According to that, the current and the previous location is employed in order to generate the trajectories of vehicles:
\begin{align*}
\Acute{x_i} &= x_i + (-s_i/\omega_i)\sin{\theta_i} + (s_i/\omega_i)\sin{(\theta_i + \omega_i\Delta{T})}, \\
\Acute{y_i} &= y_i + (s_i/\omega_i)\cos{\theta_i} +(-s_i/\omega_i)\cos{(\theta_i + \omega_i\Delta{T})}, \\
\Acute{\theta_i} &= \theta_i + \omega_i\Delta{T},
\end{align*}
where $\Delta{T}$ is the time step and $\theta_i$, $s_i$, $\omega_i$ are the heading angle, the linear and angular velocity of $i$-th vehicle, respectively. Linear and angular velocity are known also as control inputs and can be provided by on-board Inertial Measurement Unit sensor. To evaluate the effectiveness of the introduced approaches, we assume that vehicles are always members of VANET, with $\sigma_x = 3m$ and $\sigma_y = 2.5m$ resulting in average GPS localization error of $3.4m$.  Ground truth for $N=3$ vehicles and a VANET graph is shown in Fig.~\ref{traj-graph}. The simulation horizon was set to $T = 500$ time instances, with sampling interval $\Delta{T} = 0.1s$. {GPS updating time was chosen to coincide with the sampling interval $\Delta{T}$ of simulation. However, in realistic conditions updating time may be much higher, and as a matter of fact GPS's availability for real-time applications is reduced. For that purpose, the prediction step of Kalman filter can be utilized in order to provide GPS-like position \cite{Noureldin2013}, until the next true GPS measurement is provided by the sensor. In any case, instead of GPS, a visual odometry like solution which meets real-time constraints can also be used \cite{Engel2018}.} The communication range was initialized to $r_c = 20m$. We create matrices {$\boldsymbol{P^{(t)}} =  \begin{bmatrix} \boldsymbol{x^{(t)}} & \boldsymbol{y^{(t)}} \end{bmatrix} \in \mathbb{R}^{N \times 2}$} and  $\boldsymbol{W_i^{(t,k)}} = \begin{bmatrix}
\boldsymbol{w_{i,x}^{(t,k)}} & \boldsymbol{w_{i,y}^{(t,k)}}
\end{bmatrix} \in \mathbb{R}^{N \times 2}, \forall i,k$. We measured the {normalized} Average Mean Square Deviation (AMSD) \cite{Chouvardas2011} of VANET over $T$ for each iteration:
\begin{align*}
AMSD(k) = 
{\frac{1}{T}\sum_{t=1}^T\frac{1}{N}{\sum_{i=1}^N}{\frac{\norm{\boldsymbol{P^{(t)}} - \boldsymbol{W_{i}^{(t,k)}}}^2 }{\norm{\boldsymbol{P^{(t)}}}^2}}} ,
\end{align*}
and the Localization Mean Square Error (LMSE) at each time instant:
\begin{align*}
LMSE(t) = 
\frac{1}{N}{\sum_{i=1}^N}{\norm{\boldsymbol{p_i^{(t)}} - \boldsymbol{W_{i:}^{(t,K)}}}^2}
\end{align*}
Since vehicles utilizing the proposed distributed and diffusion approaches converge to the same location vector, LMSE has been computed exploiting the estimation of a random vehicle for each associated experiment. Additionally, instead of noisy GPS initialization, we utilize the coherent initialization heuristic rule of \textbf{Algorithm \ref{Init}}.
The conducted experiments were based on: i) the impact of connectivity topology of involved vehicles, i.e. VANET size $N$ and maximum number of neighbors $N_{max}$, ii) network delay effect, and iii) range measurements uncertainty. The latter dictates the feasible estimation of differential coordinates even in occluded and highly complex environments, which is considered vital for the convergence speed and location estimation accuracy of the proposed methods. Finally, we constructed the Cumulative Distribution Function (CDF) of LMSE. As it will be shown, all three proposed approaches outperform in terms of location estimation accuracy,  the distributed low cost variant \textbf{Distributed Laplacian Localization (DLL)} \cite{Piperigkos2020} of \textbf{CLL} and the method of \cite{Kim2018}, named \textbf{Maximum Likelihood based Localization (MLL)}. The former estimates only the ego-vehicle location, using the noisy positions of local neighborhood and the so-called local graph Laplacian operator. The latter is chosen since it exploits exactly the same multi-modal data as we do, while it utilizes for fusion the, prominent in CL literature \cite{Buehrer2018}, MLE. 
\subsection{Evaluation Study}
\subsubsection{Impact of VANET size}
\begin{figure*}[htbp]
  \centering
  \subfloat[VANET size $N = 3$]{\includegraphics[width=0.33\linewidth]{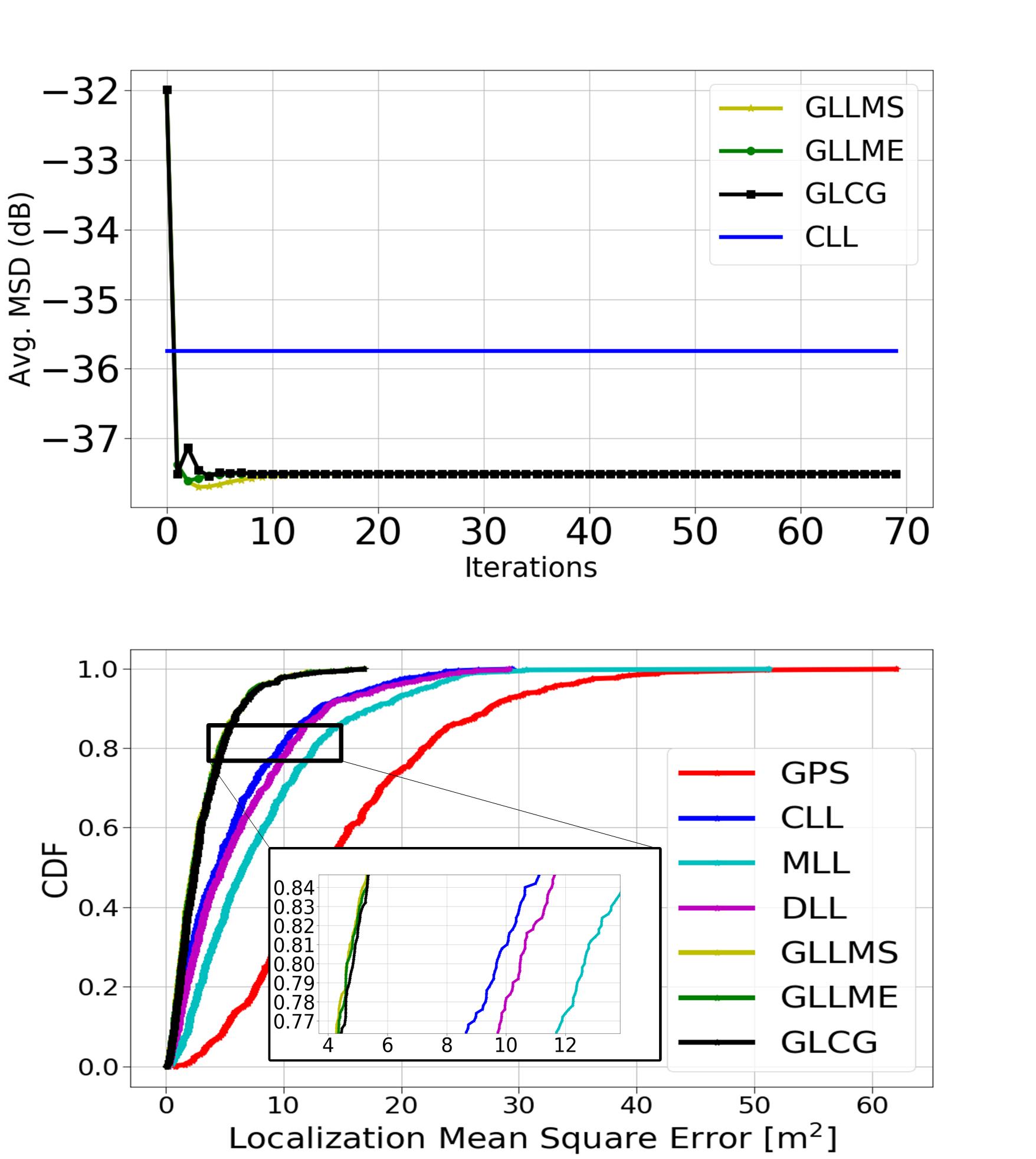}}
  \subfloat[VANET size $N = 13$]{\includegraphics[width=0.33\linewidth]{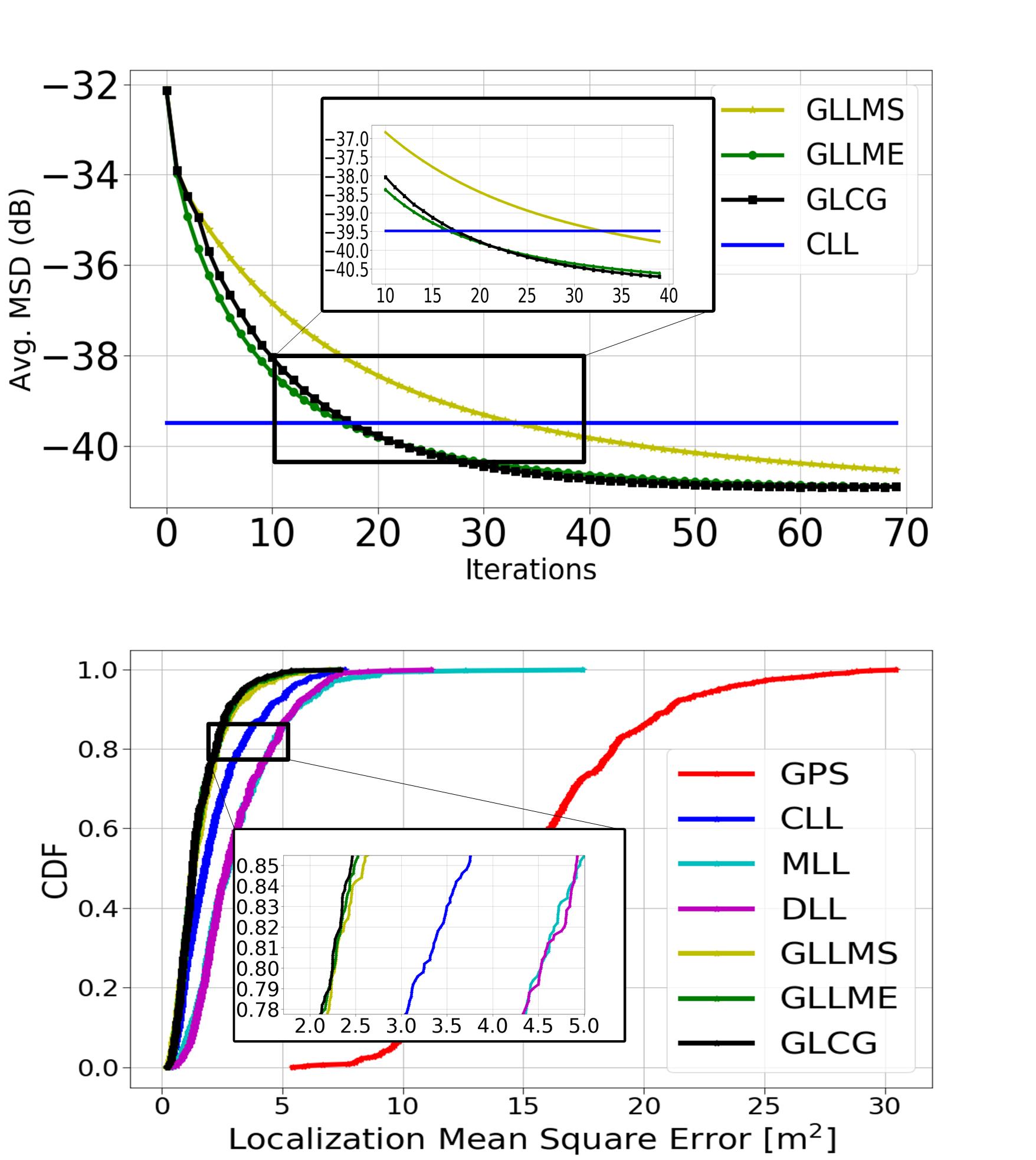}}
  \subfloat[VANET size $N = 15$]{\includegraphics[width=0.34\linewidth]{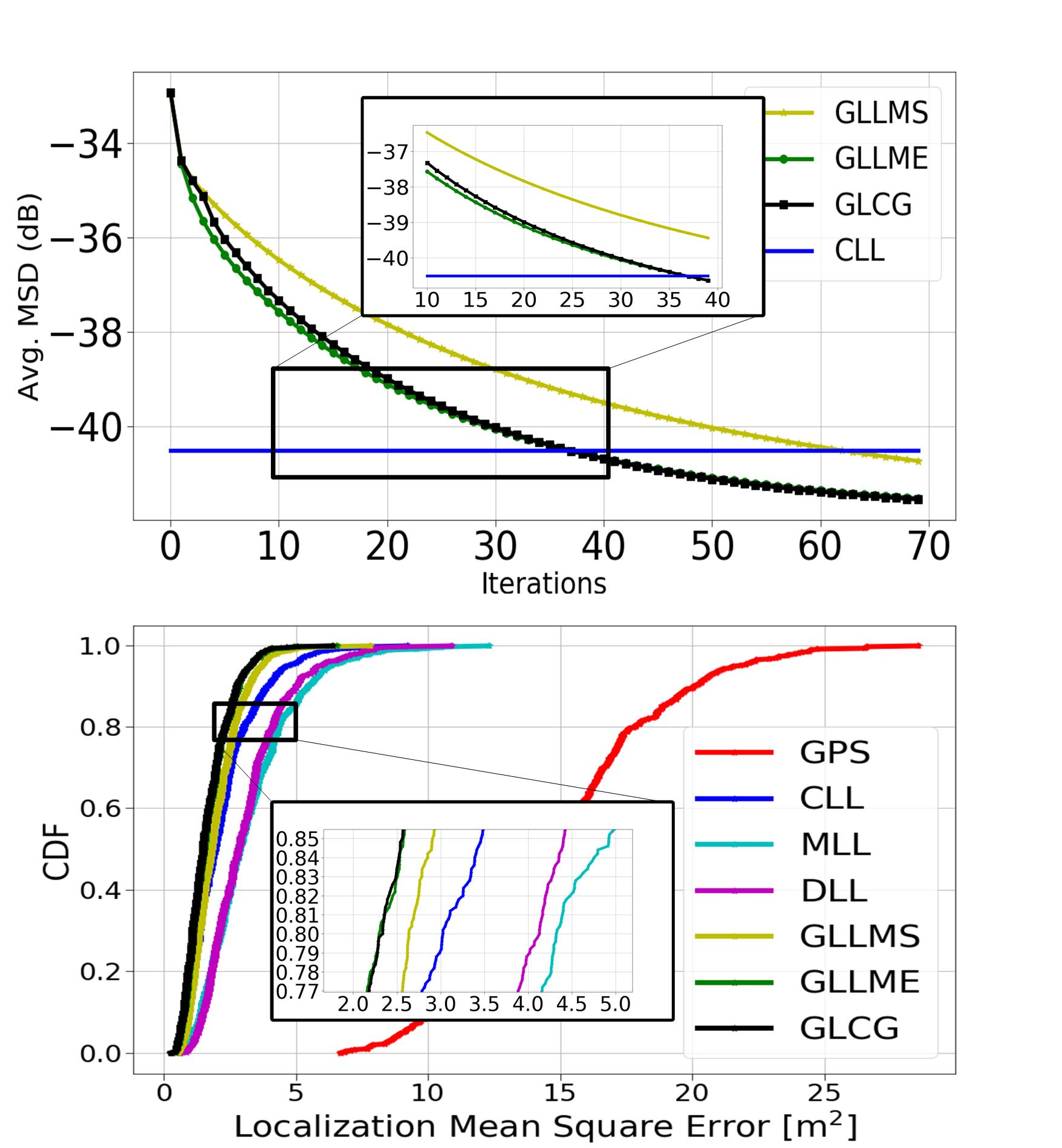}}
  \caption{Learning curves and CDF with max. neighbors $N_{max} = 6$, $\sigma_d = 1m$ and $\sigma_{az} = 4^\circ$}
  \label{vanet-size}
\end{figure*}

VANET size is a critical factor for successive graph Laplacian diffusion localization, due to the growing size of the common location vector that needs to be estimated by the vehicles. In Fig.~\ref{vanet-size}, we depict the AMSD and corresponding CDF's for $N = 3, 13$ and $15$, with $K = 70$ and $N_{max} = 6$. More specifically, in Fig.~\ref{vanet-size}-(a) all three diffusion schemes significantly outperformed \textbf{CLL}, requiring only 1 iteration, avoiding constantly receiving and broadcasting measurements. That superior performance is also present on the CDF of LMSE. The reduction of GPS LMSE is \textbf{84\%} with \textbf{GLLMS}, \textbf{GLLME} and \textbf{GLCG}, 65\% with \textbf{CLL}, 62\% with \textbf{DLL} and 53\% with \textbf{MLL}. The significant location estimation performance of the proposed approaches has been achieved since vehicles are more likely to be all-to-all V2V connected. With a larger VANET size, more iteration are required for convergence. In Fig.~\ref{vanet-size}-(b) with $N=13$, \textbf{GLLME} and \textbf{GLCG} converged faster than \textbf{GLLMS}, i.e. in around 20 iterations instead of 35. Clearly, measurements exchanges step enhanced the convergence speed of diffusion, since additional knowledge of global solution is utilized. Moreover, all three approaches outperformed all others in terms of location estimation. The reduction of GPS LMSE is \textbf{90\%} with \textbf{GLLMS}, \textbf{GLLME} and \textbf{GLCG}, 87\% with \textbf{CLL} and 80\% with \textbf{DLL} and \textbf{MLL}. Finally, in Fig.~\ref{vanet-size}-(c) with $N = 15$, \textbf{GLLME} and \textbf{GLCG} converged in around 40 iterations, instead of 60 required by \textbf{GLLMS}. Moreover, the reduction of GPS LMSE is \textbf{87\%} with \textbf{GLLMS}, \textbf{89\%} with \textbf{GLLME} and \textbf{GLCG}, 86\% with \textbf{CLL} and 80\% with \textbf{DLL} and \textbf{MLL}. Evidently, \textbf{GLLME} and \textbf{GLCG} are much more efficient in terms of convergence speed than \textbf{GLLMS}, when VANET size grows. At the same time, all three methods exhibited superior location estimation performance over \textbf{CLL}, \textbf{DLL} and \textbf{MLL}.

\begin{figure*}[htbp]
  \centering
  \subfloat[\textbf{GLLMS} convergence]{\includegraphics[width=0.33\linewidth]{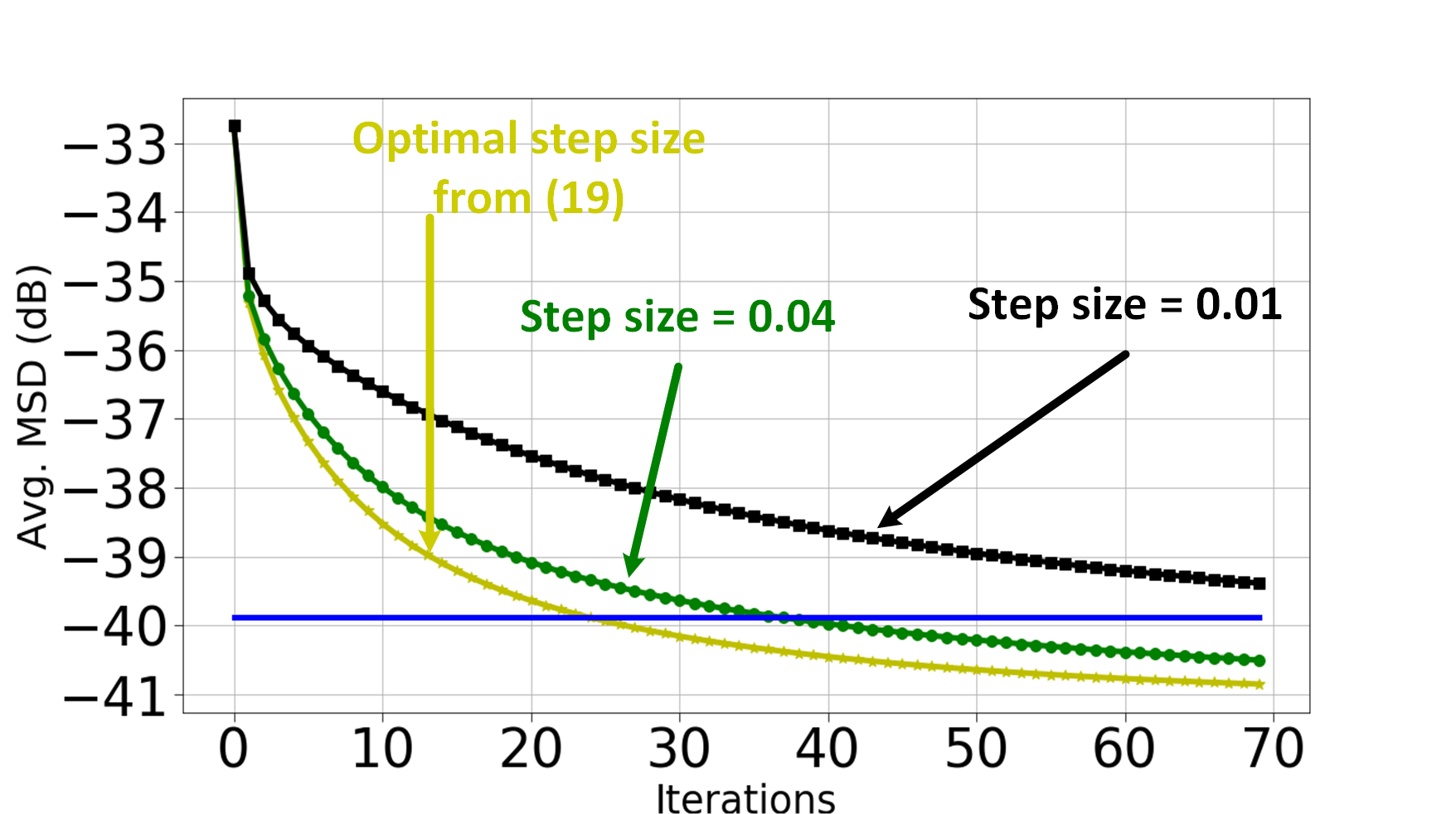}}
  \subfloat[\textbf{GLLME} convergence]{\includegraphics[width=0.33\linewidth]{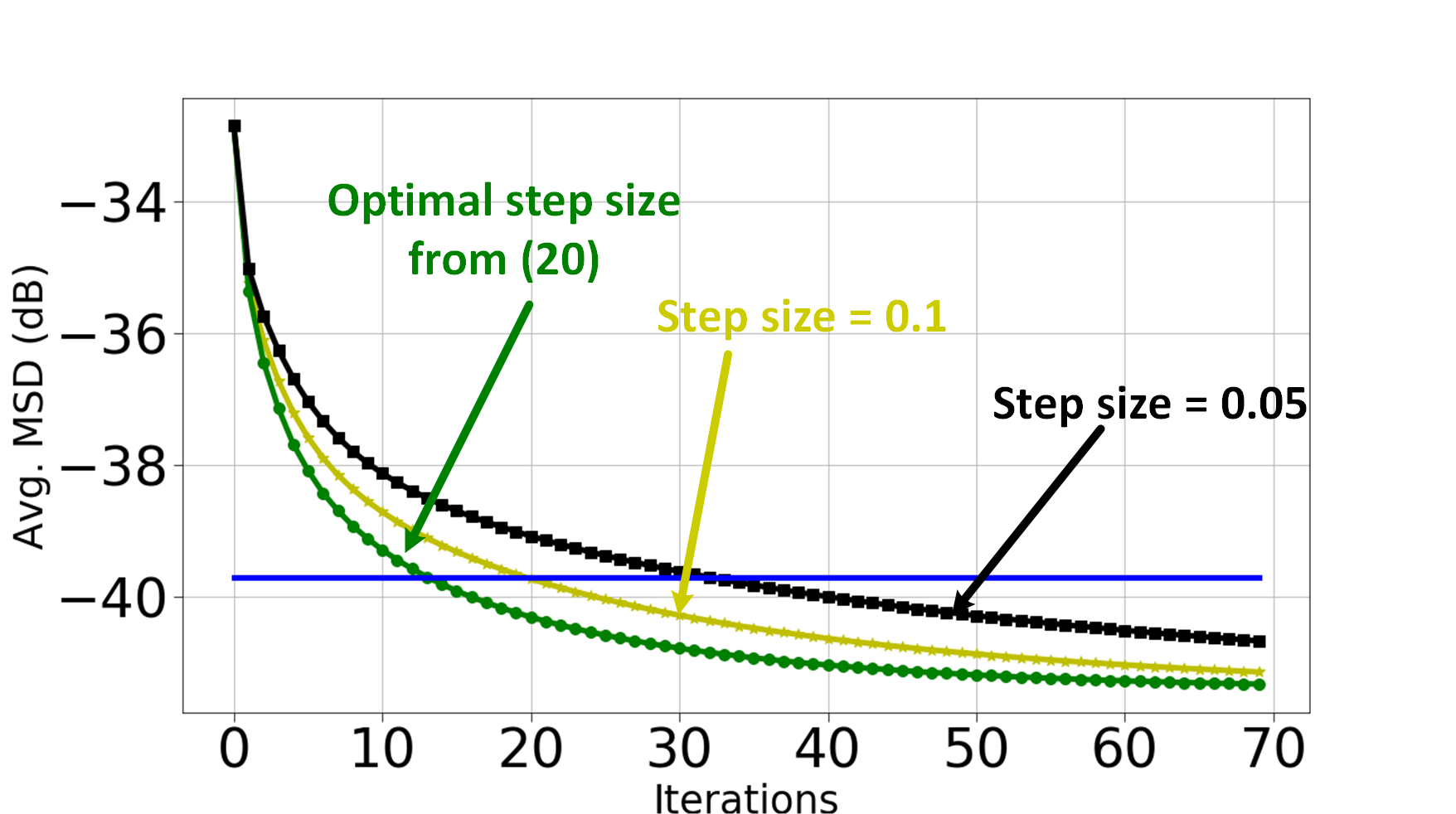}}
  \subfloat[\textbf{GLLME} sufficient condition for step size]{\includegraphics[width=0.33\linewidth]{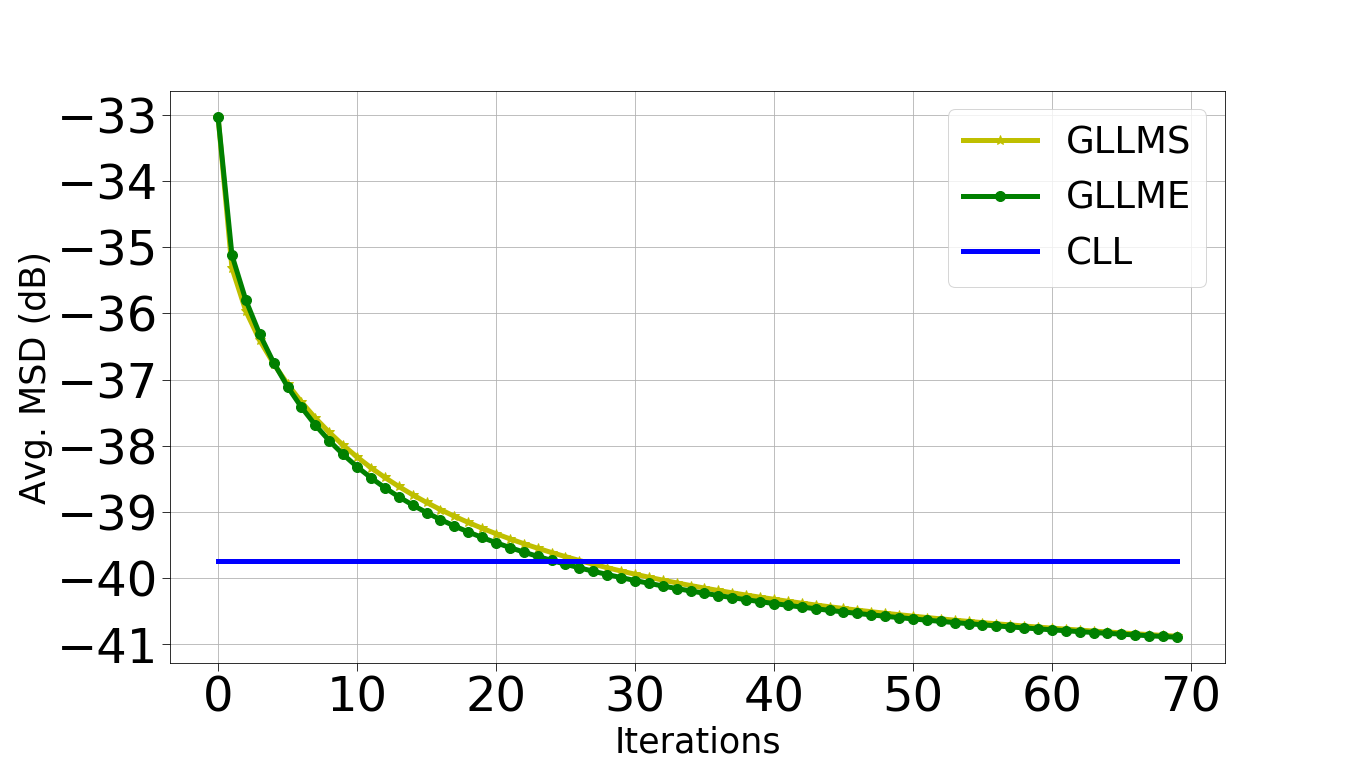}}
  \caption{Step size selection for VANET size $N = 10$, max. neighbors $N_{max} = 6$, $\sigma_d = 1m$ and $\sigma_{az} = 4^\circ$}
  \label{step-size}
\end{figure*}

Step size selection influences the convergence speed of both \textbf{GLLMS} and \textbf{GLLME}, as demonstrated in Fig.~\ref{step-size}. More specifically, in Fig.~\ref{step-size}-(a), \textbf{GLLMS} with step sizes close to zero, i.e. $0.04$ and $0.01$, resulted in higher number of required iterations. \textbf{GLLMS} with optimal step size according to \eqref{m1}, achieved the fastest convergence of around 20 iterations. The same behavior is also present with \textbf{GLLME} in Fig.~\ref{step-size}-(b). Step sizes $0.1$ and $0.05$ required almost 20 and 30 iterations, while the optimal selection \eqref{m2} resulted in the fastest convergence ($\sim$ 10 iterations). Finally, we depict in Fig.~\ref{step-size}-(c) the learning curves of both \textbf{GLLMS} and \textbf{GLLME}, the latter with step size according to unbiasedeness sufficient condition. Clearly, \textbf{GLLME} converges with much slower speed, almost identical to that of \textbf{GLLMS}. Thus, the benefits of measurements exchanges are now suppressed.

\subsubsection{Impact of vehicle connections}
\begin{figure*}[htbp]

  \centering
  \subfloat[\textbf{GLCG} convergence ]{\includegraphics[width=0.33\linewidth]{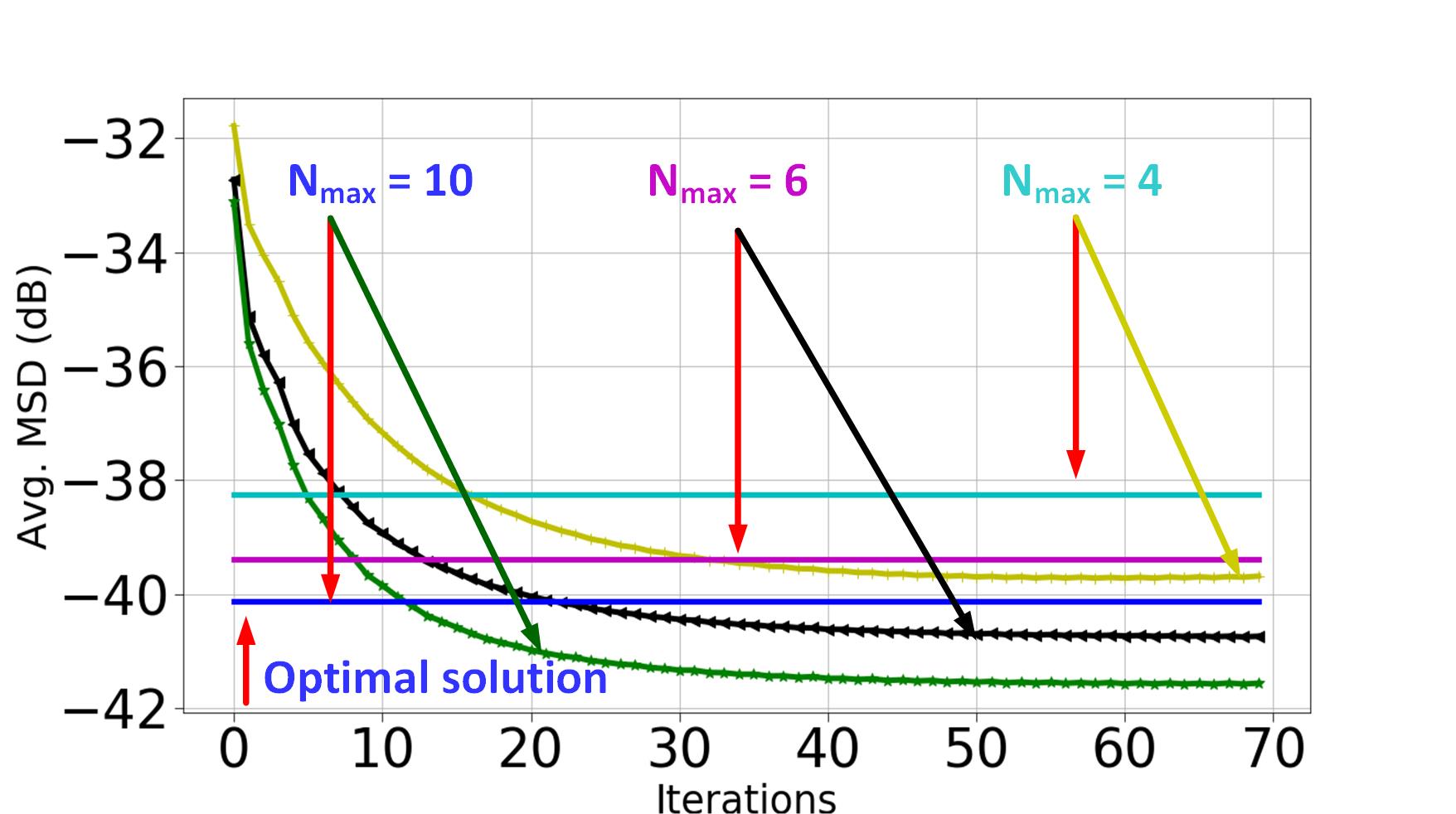}}
  \subfloat[CDF with max. neighbors $N_{max}=4$ ]{\includegraphics[width=0.33\linewidth]{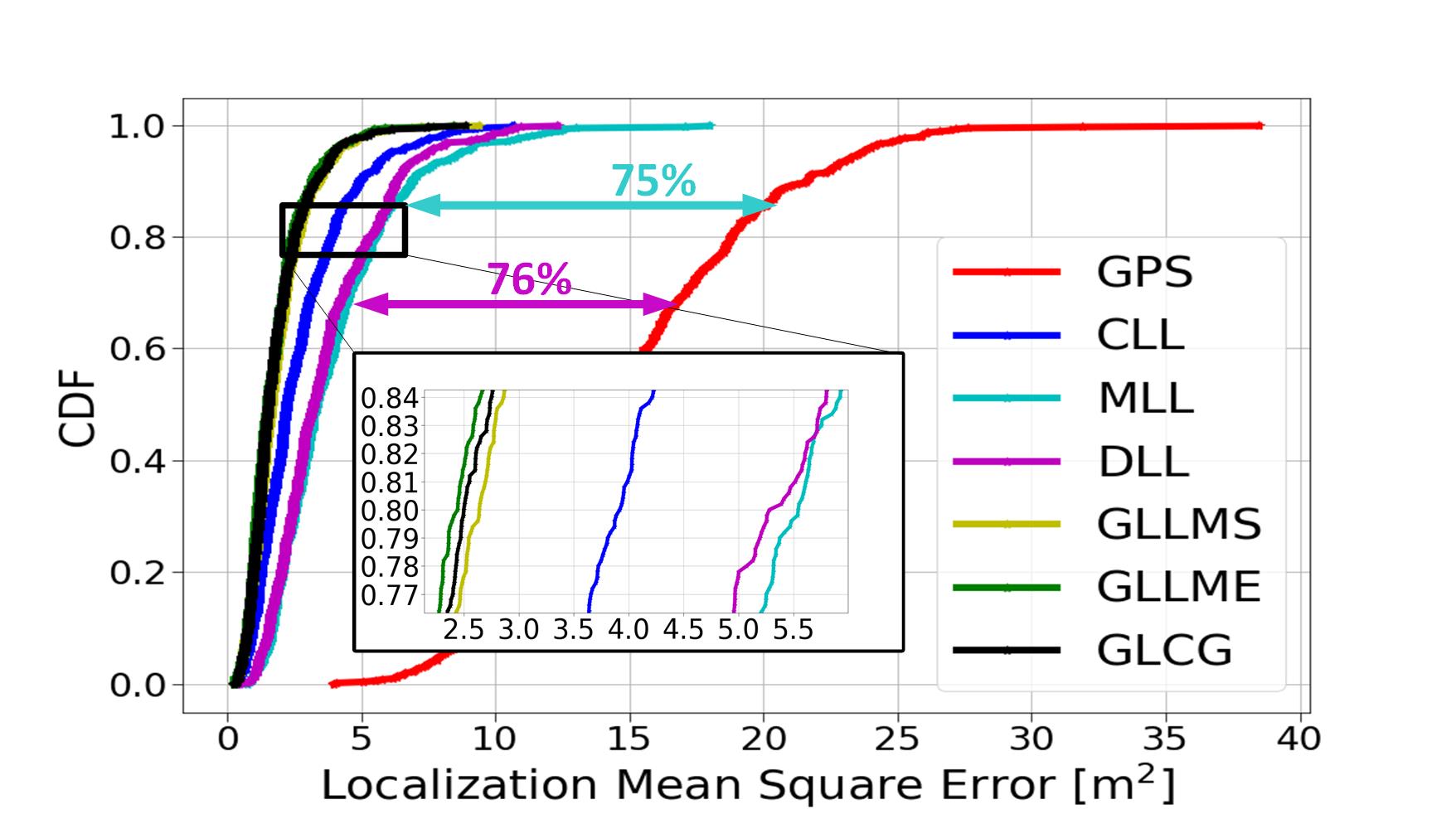}}
  \subfloat[CDF with max. neighbors $N_{max}=10$ ]{\includegraphics[width=0.33\linewidth]{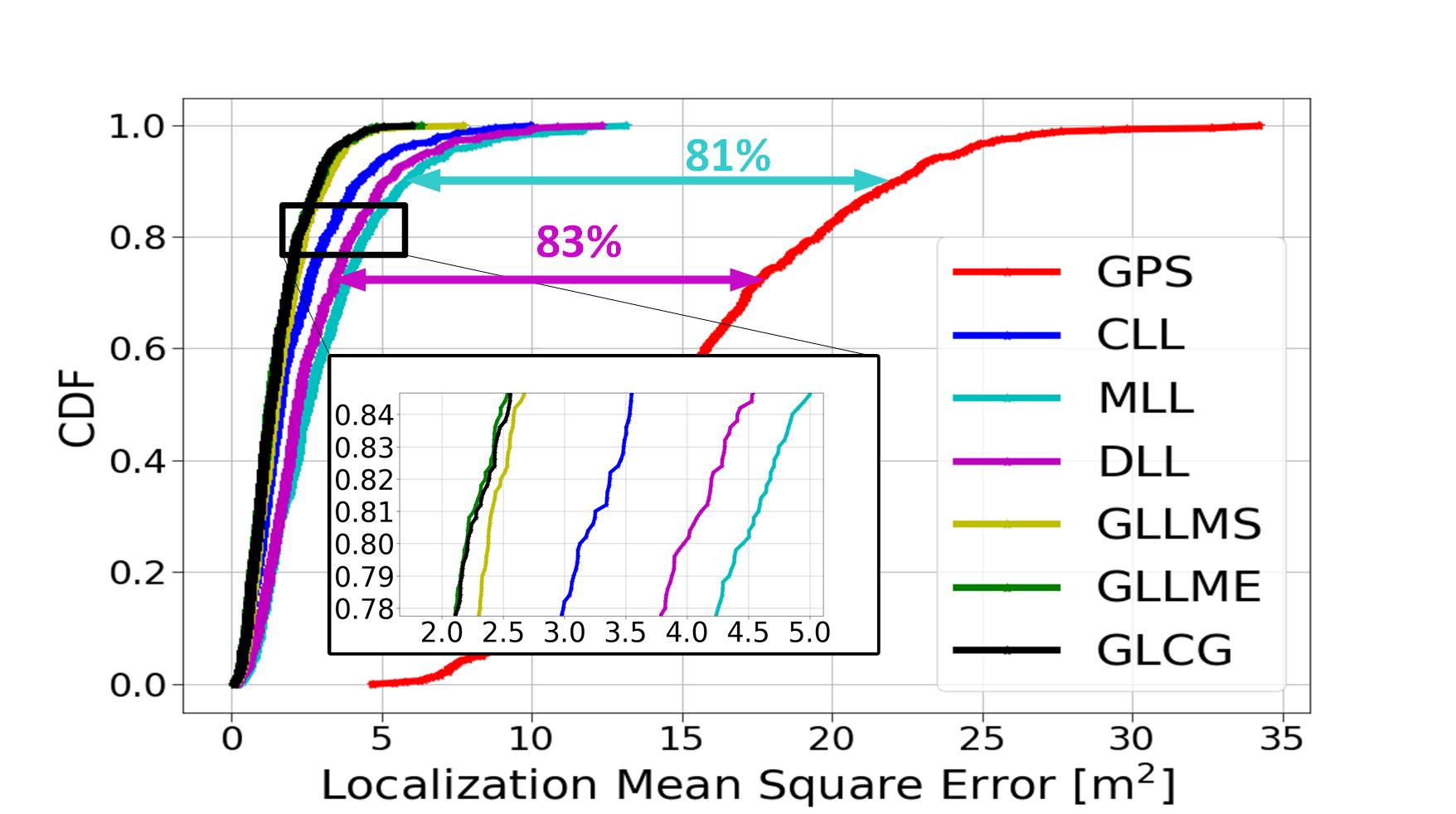}}
  \caption{Impact of connections with VANET size $N = 10$, $\sigma_d = 1m$ and $\sigma_{az} = 4^\circ$}
  \label{connections}
\end{figure*}

Vehicles may communicate with a different number of neighbors while they are moving. Since the graph Laplacian regression vectors represent in fact those possible V2V connections, it is straightforward to study the impact of $N_{max}$ to the diffusion schemes. Therefore, the effect of that parameter in convergence speed and location estimation accuracy is demonstrated in Fig.~\ref{connections}, with $N = 10$ and $K = 70$. For example, in Fig.~\ref{connections}-(a) we depict the AMSD of \textbf{GLCG} for $N_{max} = 4, 6$ and $10$. Clearly, the optimal \textbf{CLL} solution is achieved when $N_{max} = 10$, since vehicles integrate greater amount of information. Accordingly, \textbf{GLCG} with $N_{max} = 10$ converged much faster to the optimal solution, i.e. in around 10 iterations. Each vehicle is connected to a large number of neighbors, even to the overall number of VANET's vehicles, and thus it is much for efficient to estimate the entire location vector in a fully distributed manner. Additionally, \textbf{GLCG} with $N_{max} = 6$ converges in around 20 iterations, while $N_{max} = 4$ requires much more iterations than $K = 70$. In Fig.~\ref{connections}-(b), the reduction of GPS LMSE is \textbf{88\%} with \textbf{GLLMS}, \textbf{GLLME} and \textbf{GLCG} and 83\% with \textbf{CLL}. In Fig.~\ref{connections}-(c), with $N_{max} = 10$, performances have been improved. For example, the reduction of GPS LMSE is \textbf{90\%} with \textbf{GLLMS}, \textbf{GLLME} and \textbf{GLCG} and 87\% with \textbf{CLL}. Once again, the proposed distributed and diffusion schemes outperformed not only the global solution, but also \textbf{DLL} and \textbf{MLL}. Consequently, the impact of V2V connections, in the form of graph Laplacian regression vectors, is crucial for the convergence speed. 

{As a final experiment, we measured the algebraic connectivity of VANET graph with size $N = 13$, maximum number of neighbors $N_{max} = 6$ and $\sigma_d = 1m, \sigma_a = 4^\circ$. Algebraic connectivity \cite{Li2017} equals to the second smallest eigenvalue of Laplacian matrix (of VANET graph) and indicates the time-varying topology of VANET. Lower (higher) connectivity implies weakly (strongly) connected graphs. From Fig.~\ref{connectivity-impact}-(a), it is derived that connectivity values range between 0.3 and 0.8 resulting in a rather weakly connected graph. The corresponding LMSE attained by \textbf{GLLMS} with $K = 70$ is demonstrated in Fig.~\ref{connectivity-impact}-(b) (magenda color). In addition, the same VANET was created assuming that all vehicles communicate with all others (All to All connection), and the resultant LMSE is shown with red color. In that case, superior performance has been attained, i.e. maximum LMSE equal to $\sim 6 m^2$ instead of $8 m^2$. Therefore, with strongly connected VANET graphs (i.e. vehicles with many V2V neighbors), it is expected that localization accuracy can be further increased. However, restricted communication capabilities or allowed number of neighbors may be prohibitive for forming those type of VANET graphs.
\begin{figure}[htbp]
  \centering
  \subfloat[Connectivity of VANET graph]{\includegraphics[width=0.5\linewidth]{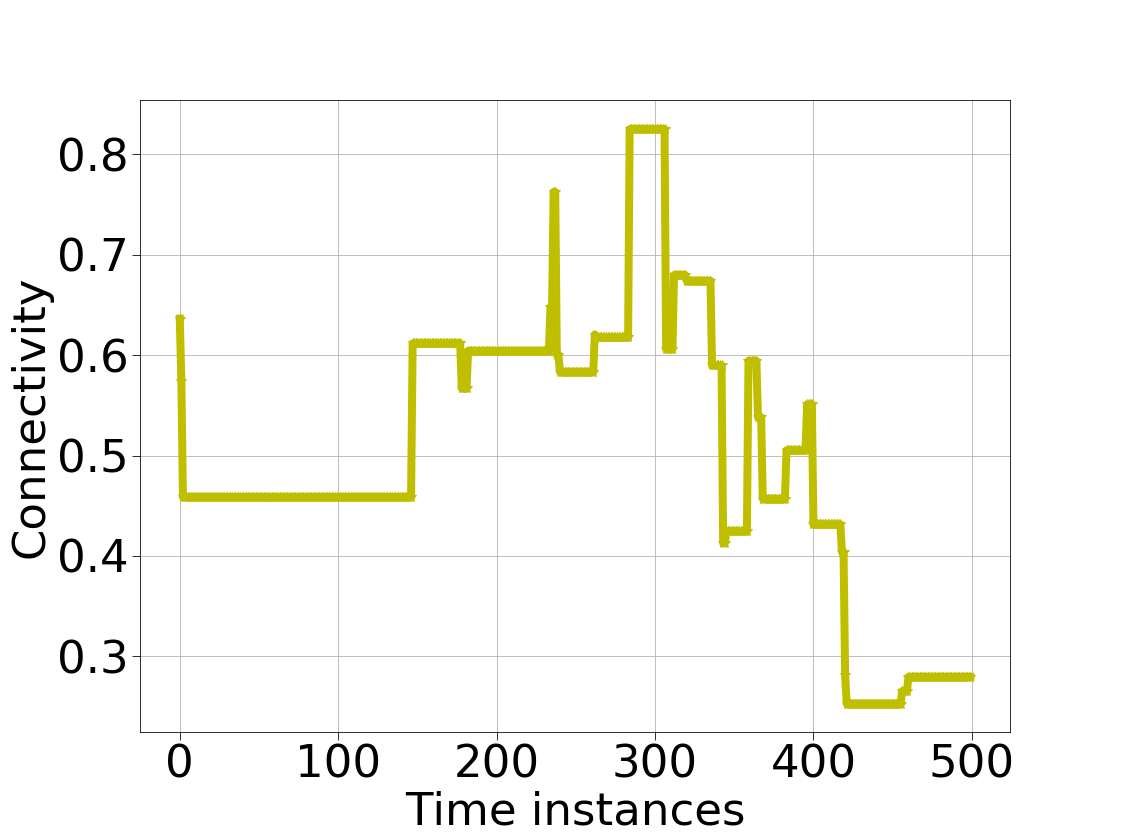}}
  \subfloat[\textbf{GLLMS} performance]{\includegraphics[width=0.5\linewidth]{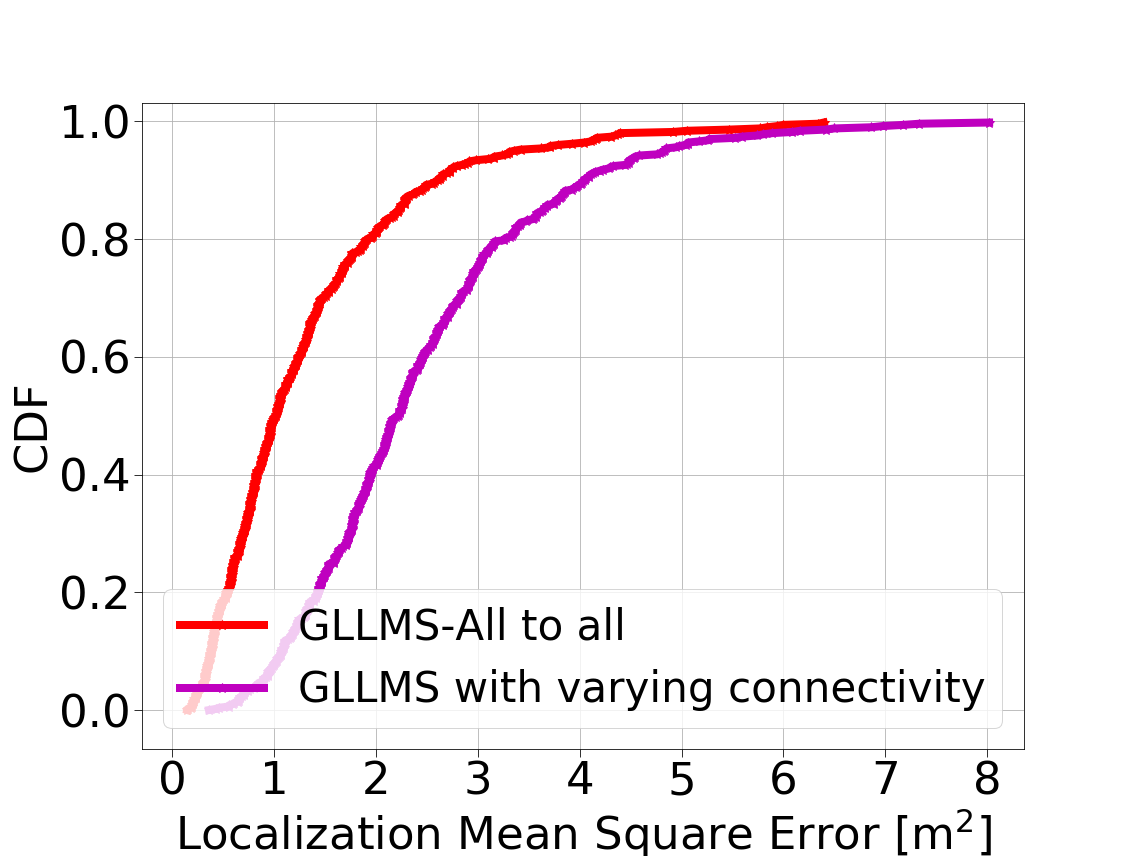}}
  \caption{Impact of weakly and strongly connectivity}
  \label{connectivity-impact}
\end{figure}
}
\subsubsection{Impact of network delay}
The upcoming 5G standard will include Ultra Reliable Low Latency  Communication (URLLC) services. URLLC is designed for applications that require stringent latency and reliability requirements in vehicular communications \cite{Ge2019}. Clearly, the network of vehicles must be time synchronized, which means that each vehicle transmits to neighbors the intermediate vectors before the next iteration $k+1$. However, according to URLLC specifications, communication  delay  can  be  regarded  around $10ms$ for a network of approximately $10$ vehicles, with velocities  lower than $10m/s$. To this end, and based on \cite{Hua2020}, the location estimation vector at every iteration is now provided by: $\boldsymbol{w_{i,x}^{(t,k+1)}} = \sum_{l \in \mathcal{N}_i}c_{il}^{(t)}\boldsymbol{\psi_{l,x}^{(t,k+1-\tau)}}$. Small integer $\tau > 0$ indicates the delayed version of intermediate vectors received by vehicle $i$. {Each vehicle broadcasts CAM messages at least every 100$ms$ \cite{2018}, while the maximum delay introduced by V2V communication can reach 300$ms$ \cite{Yao2013} at heavy traffic density 0.1 vehicles/meter. Therefore, for every iteration of proposed algorithms we have at most 400 $ms$ delay, which imply that $\tau$ can reach 4, i.e. vehicle $i$ receives vectors $\boldsymbol{\psi_{l,x}^{(t, k+1-4)}}$ by its neighbors, estimated 4 iterations before.}

\begin{figure*}[htbp]
  \centering
  \subfloat[$\tau = \begin{bmatrix} 1, & 2, & 3, & 4 \end{bmatrix}$ ]{\includegraphics[width=0.33\linewidth]{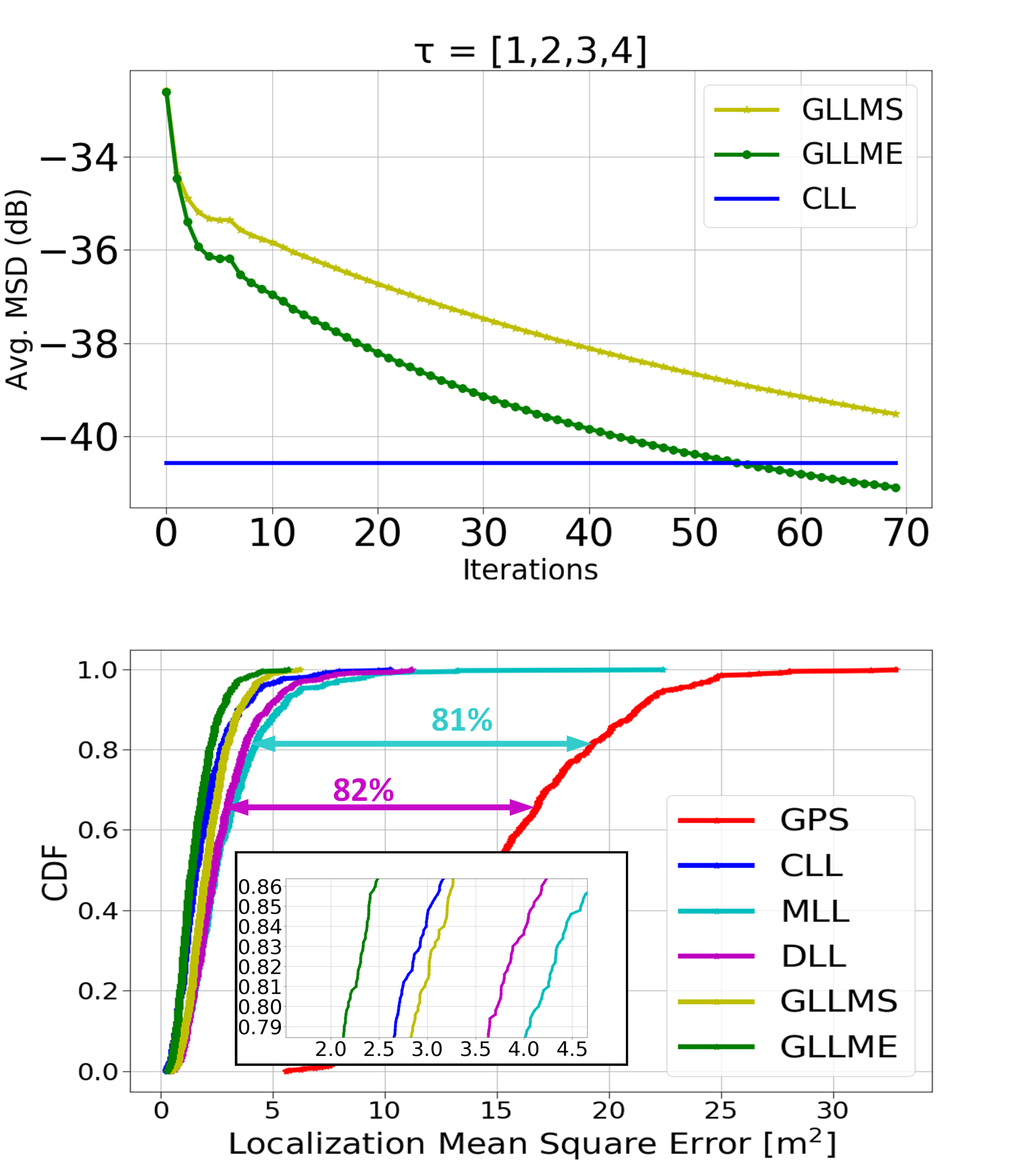}}
  \subfloat[$\tau = 4$ for 80\% of neighbors]{\includegraphics[width=0.33\linewidth]{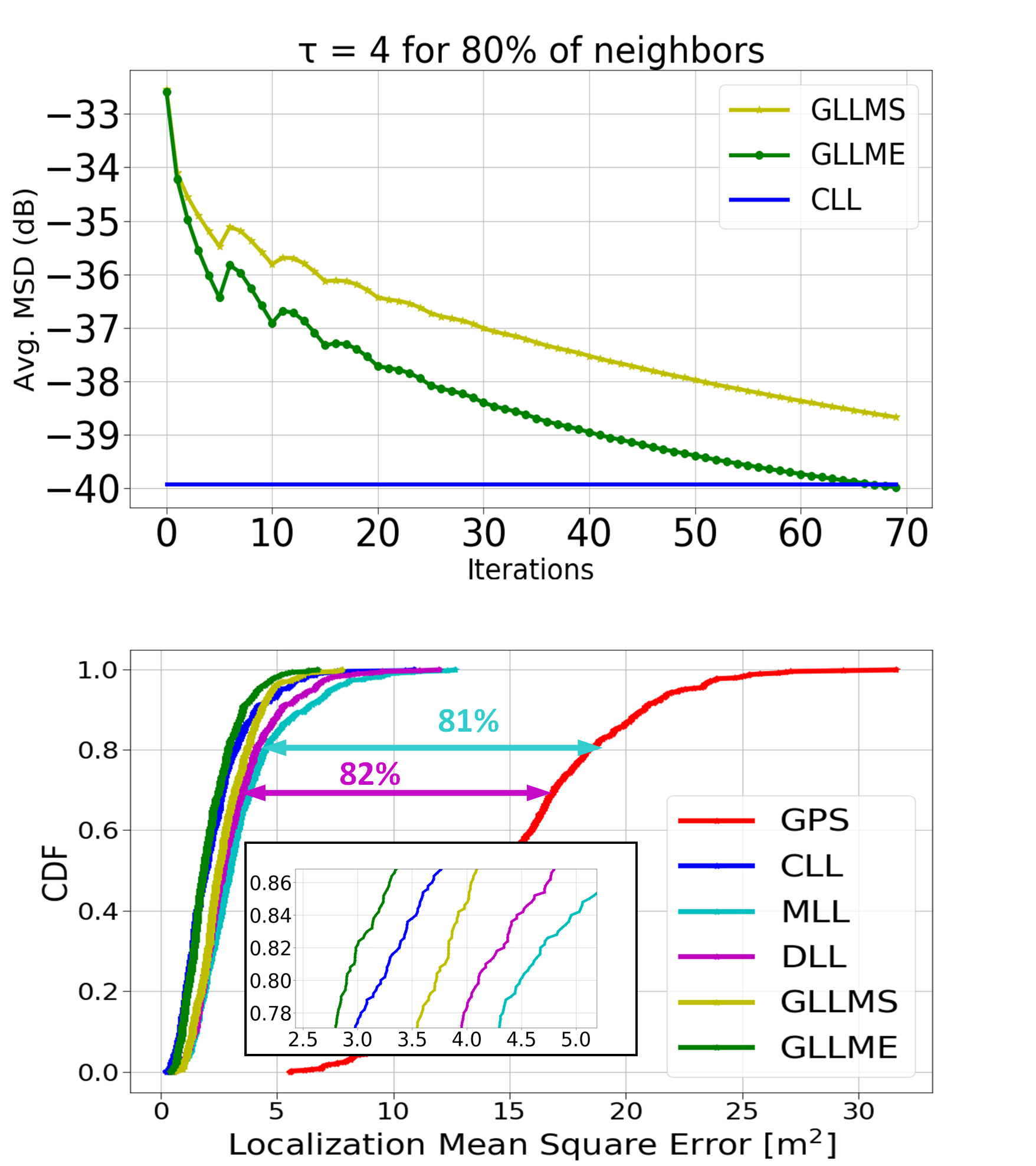}}
  \subfloat[$\tau = 4$ for 100\% of neighbors]{\includegraphics[width=0.33\linewidth]{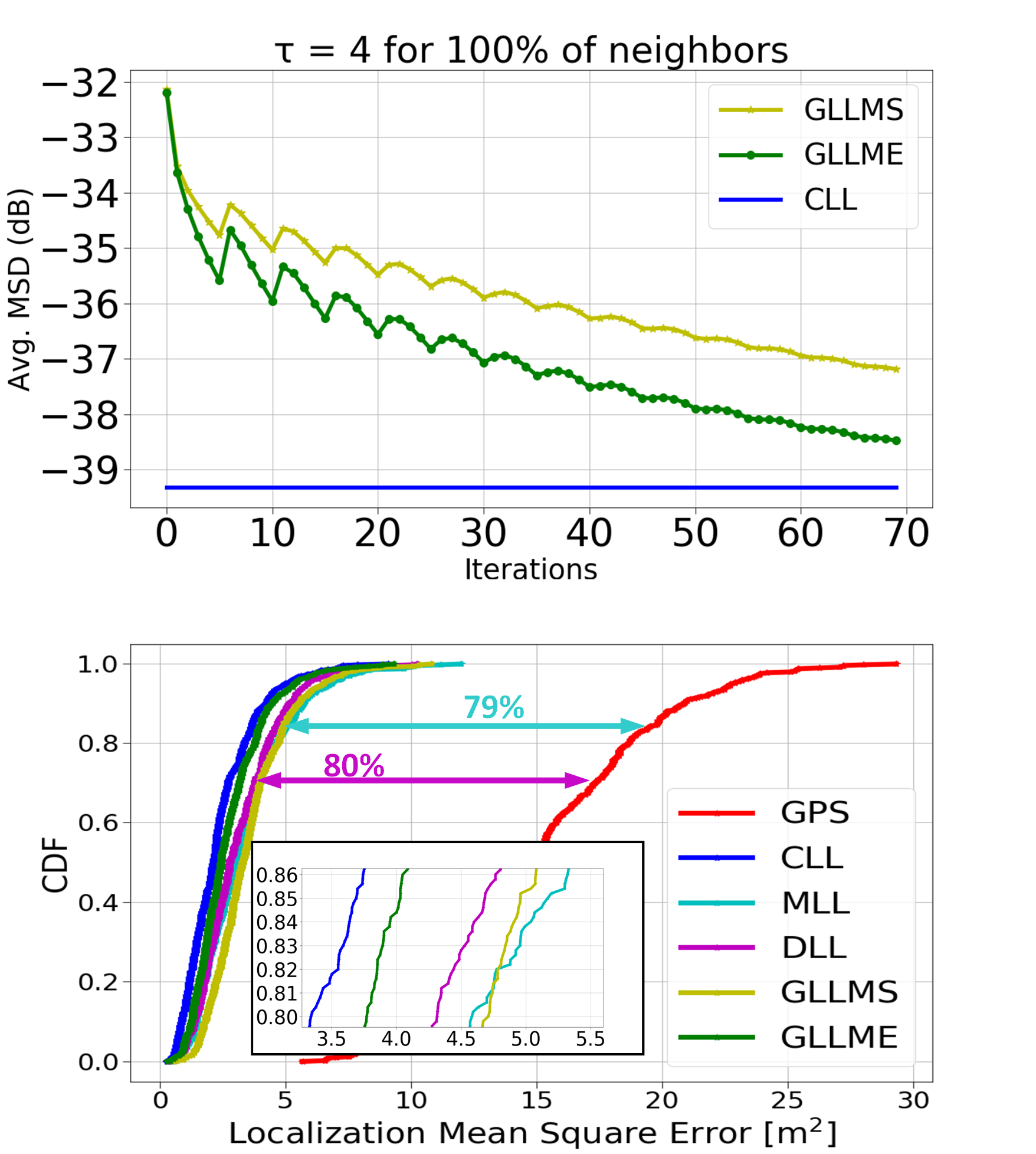}}
  \caption{{Network delay study with VANET size $N = 13$, max. neighbors $N_{max} = 6$, $\sigma_d = 1m$ and $\sigma_{az} = 4^\circ$}}
  \label{net_delay}
\end{figure*}

Network delay effect in graph Laplacian diffusion is demonstrated in Fig.~\ref{net_delay}, with $N = 13$ and $K = 70$. The \textbf{GLCG} has been omitted from evaluation, since it was verified during the conducted experiments that totally failed to operate. That drawback is related to computing the optimal step size $a_i$ and factor $\beta_i$ directly from available data. Due to the fact that past estimations are actually used, the two main parameters are significantly deviate from their expected values and negatively influence the convergence of \textbf{GLCG}. 
The \say{bad} condition number $\kappa(\boldsymbol{U_i^{(t)}})^2$ additionally affects \textbf{GLCG}.
{In Fig.~\ref{net_delay}-(a), the network delay parameter, for each vehicle and iteration number, is randomly chosen (with
    probability 0.25) to take values from the set $\tau = \begin{bmatrix} 1, & 2, & 3, & 4 \end{bmatrix}$ in order to simulate random time-varying delay for each vehicle. Clearly, \textbf{GLLME} converges much faster than \textbf{GLLMS}. The latter actually requires more iterations than 70 in order to converge effectively. However, the convergence speed in general, has been significantly reduced compared to Fig.~\ref{vanet-size}-(b). \textbf{GLLME} achieved \textbf{90\%} reduction of LMSE (greater than \textbf{CLL}), while \textbf{GLLMS} \textbf{86\%}. To further investigate the impact of delay, we set that each vehicle receives from 80\% of its neighbors the intermediate vectors estimated $\tau = 4$ iterations before. From Fig.~\ref{net_delay}-(b), we derive that \textbf{GLLME} attains much higher convergence speed than \textbf{GLLMS} once again, though significantly increased with respect to previous case. \textbf{GLLME} achieved \textbf{86\%} reduction of LMSE, while \textbf{GLLMS} \textbf{82\%}. Finally, with $\tau = 4$ for all vehicles and their neighbors, network delay has a much stronger footprint, as demonstrated in  Fig.~\ref{net_delay}-(c). Once again \textbf{GLLME} exhibits higher convergence speed, though it can't converge during $K = 70$ iterations. Network delay impacts also on location estimation accuracy, since  \textbf{GLLME} achieved \textbf{83\%} reduction of LMSE, while \textbf{GLLMS} \textbf{77\%}, both lower than \textbf{CLL}. Therefore, we conclude that \textbf{GLLME} seems to be more robust to network delay effect, i.e. time-varying delay parameters and constant parameter with stronger footprint, than \textbf{GLLMS} both for the case of convergence and location accuracy.}

\subsubsection{Impact of range measurements noise}
\begin{figure*}[htbp]
  \centering
  \subfloat[\textbf{GLLME} convergence ]{\includegraphics[width=0.33\linewidth]{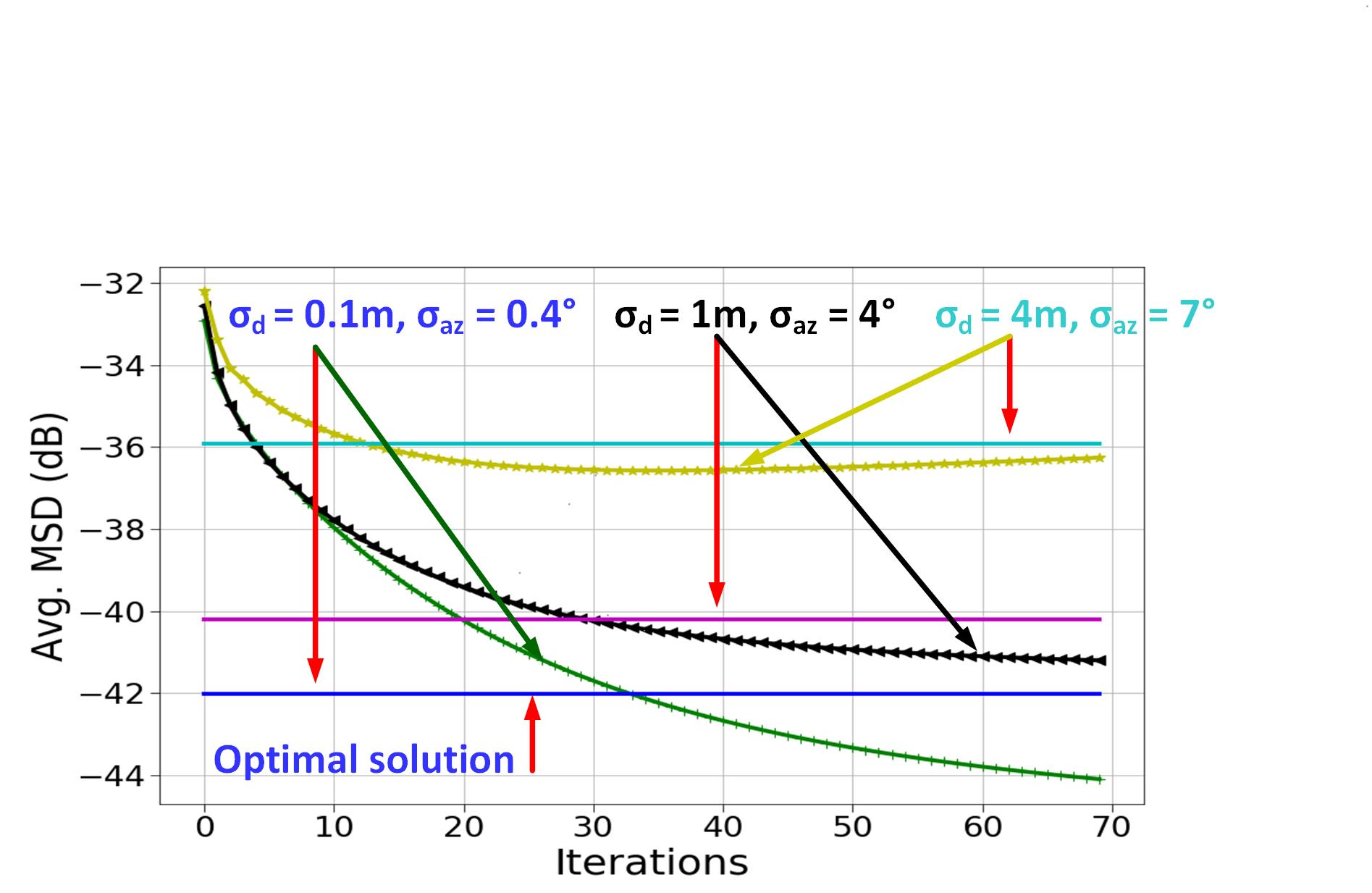}}
  \subfloat[CDF with $\sigma_d = 0.1m, \sigma_{az} =0.4^\circ$ ]{\includegraphics[width=0.33\linewidth]{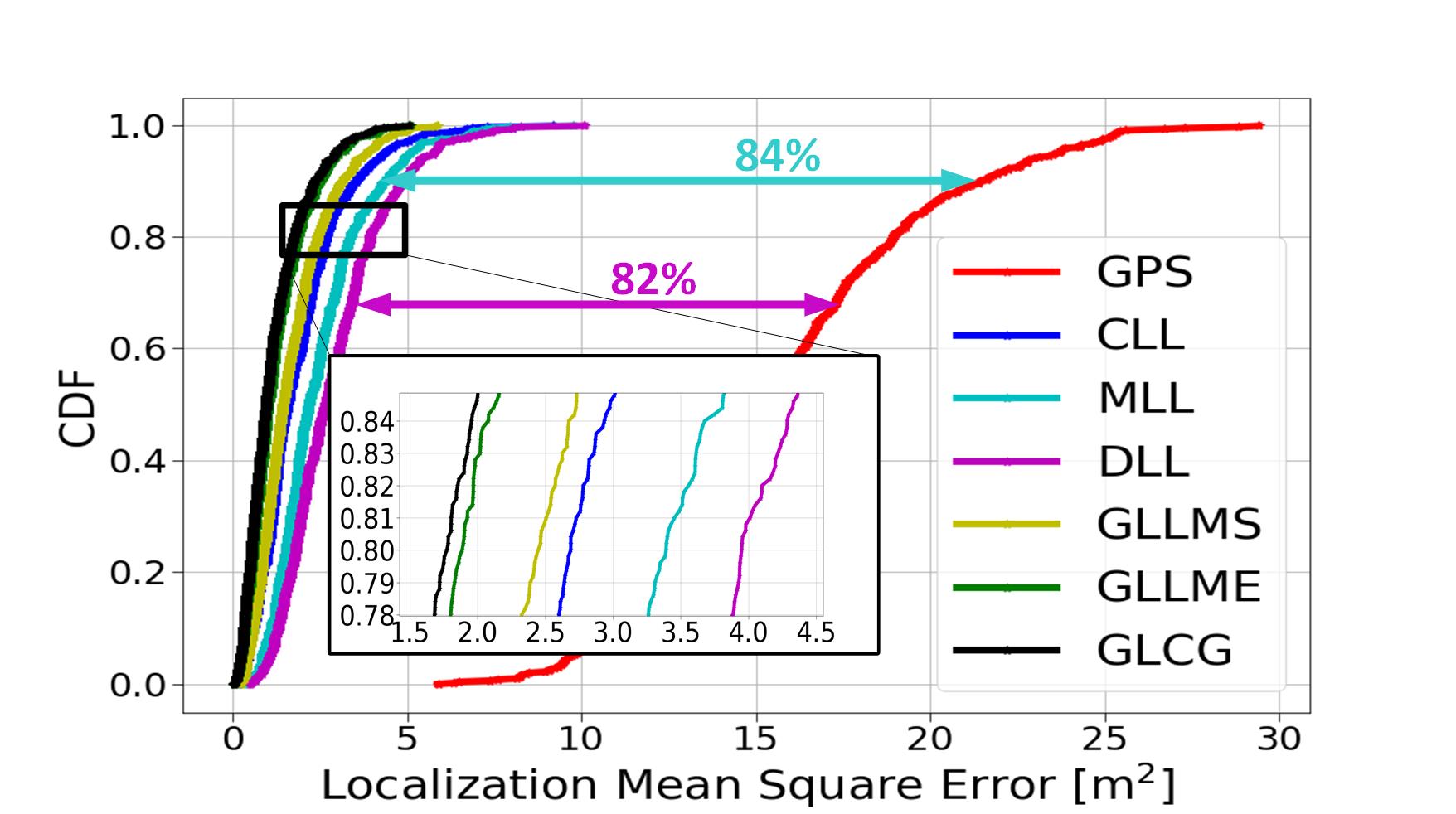}}
  \subfloat[CDF with $\sigma_d = 4m, \sigma_{az} = 7^\circ$ ]{\includegraphics[width=0.33\linewidth]{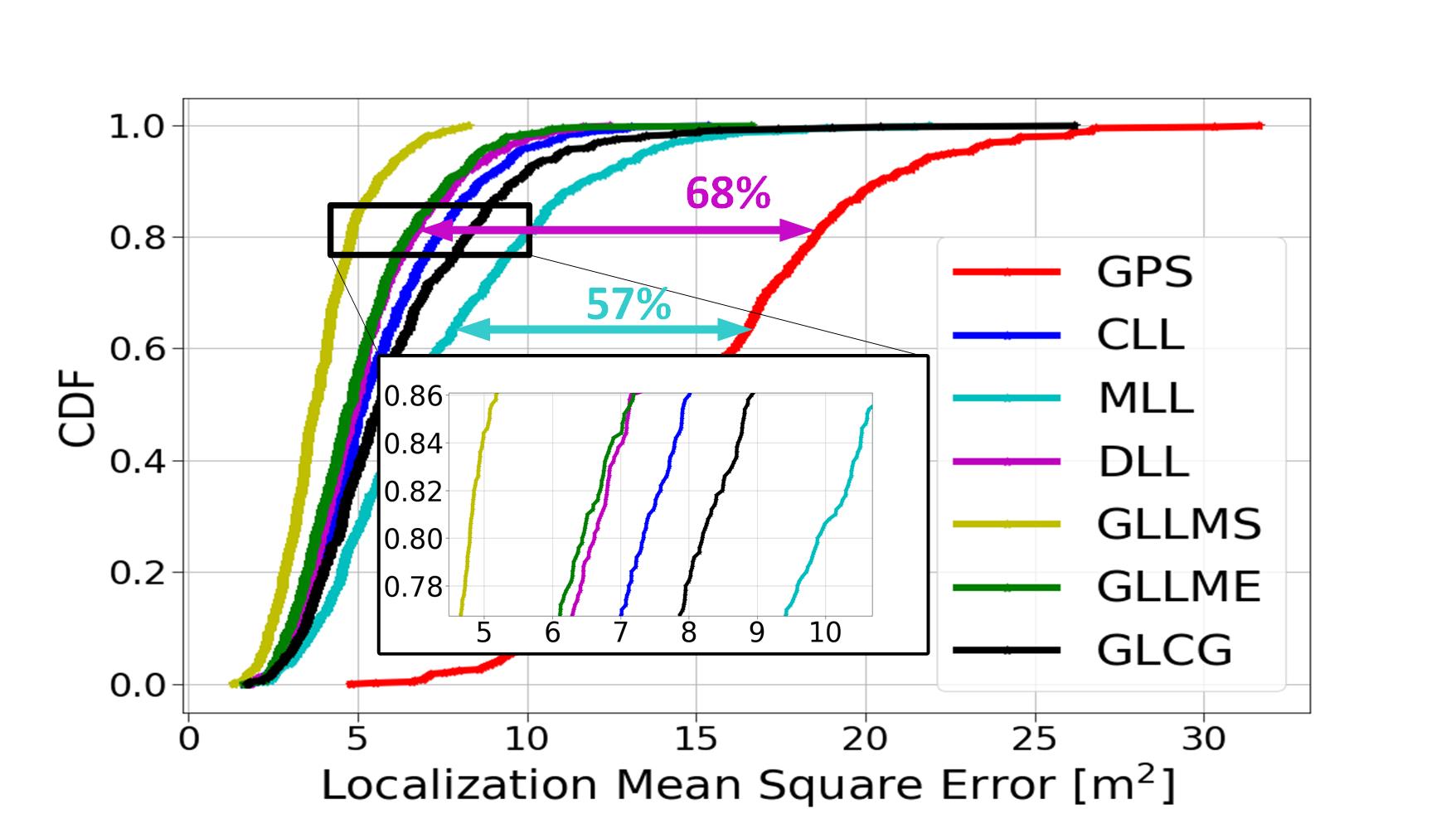}}
  \caption{Range measurements uncertainty study, assuming a VANET with $N = 15$ vehicles, while the maximum number of neighbors is $N_{max} = 6$}
  \label{fig-noise}
\end{figure*}

In highly complex urban environments, vehicles may exhibit non-line-of-sight conditions, limiting their ability of accurate relative measurements. The impact of uncertainty in range measurements is depicted in Fig. ~\ref{fig-noise}, with $N = 15$ and $K = 70$. Differential coordinates with increased measurement noise may deteriorate the performance of diffusion schemes. For example, in Fig. ~\ref{fig-noise}-(a) we present the AMSD for \textbf{GLLME}. The optimal \textbf{CLL} solution has been achieved with the lowest range noise, i.e. $\sigma_d = 0.2m$ and $\sigma_{az} = 0.4^\circ$, since it facilitates the feasible estimation of differential coordinates. The \textbf{GLLME} converges in around 30 iterations. However, when range noise increases, \textbf{GLLME} fails to converge during $K = 70$ iterations. Especially in the case of $\sigma_d = 4m$ and $\sigma_{az} = 7^\circ$, \textbf{GLLME} is far from the optimal global solution. In Fig.~\ref{fig-noise}-(b), the reduction of GPS LMSE is \textbf{89\%} with \textbf{GLLMS}, \textbf{92\%} with \textbf{GLLME} and \textbf{GLCG} and 88\% with \textbf{CLL}. Evidently, \textbf{GLLME} and \textbf{GLCG} outperformed all others, including \textbf{DLL} and \textbf{MLL}. However, in Fig.~\ref{fig-noise}-(c), the reduction of GPS LMSE is \textbf{76\%} with \textbf{GLLMS}, \textbf{69\%} with \textbf{GLLME}, \textbf{64\%} with \textbf{GLCG} and 66\% with \textbf{CLL}. The \textbf{GLCG} approach exhibits lower performance than all others, except \textbf{MLL}. The \textbf{GLLMS} approach proves its robustness, since each vehicle utilizes only its own differential coordinate, and thus limiting their noisy impact. The two other diffusion schemes are severely degraded, due to their main feature: receiving the noisy differentials of their connected neighbors. Thus, range measurements uncertainty is a critical factor of both convergence and location estimation of graph Laplacian diffusion.

Therefore, we conclude that measurements exchanges step significantly increases the convergence speed towards the global solution of \textbf{CLL}, by broadcasting only a scalar value and a sparse vector. At the same time, LMS with measurements exchanges exhibits almost the same convergence speed and location estimation accuracy with CG, and even outperform it in the case of network delay and increased range measurements noise. Thus, it isn't required to optimally select the step size as the optimizing argument of the gradient, or to perform conjugate steps towards the solution as \textbf{GLCG} does. An LMS based solution, exploiting step sizes according to \textbf{Proposition 1} and \textbf{2}, would be suffices for effective graph Laplacian diffusion localization.  
\subsubsection{Execution time}
Vehicles utilizing the proposed approaches should constantly broadcast and receive information for $K$ iterations, in order to converge to the optimal solution. Furthermore, although measurements exchanges step increased convergence speed, it is expected to increase time complexity, due to the additional exploitation of $\{\delta_l^{(t,x)}, \boldsymbol{L_{l:}^{(t)T}} \}$ pair. {Instead of iteration time interval for each one of the vehicles, we measured the total average execution time of the proposed diffusion schemes over $T = 500$ and demonstrated it below in TABLE \ref{tab2}, with maximum number of iterations $K = 70$, $N_{max} = 6, \sigma_d = 1m$ and $\sigma_{az} =4^\circ$.
All three methods seem to be effective in critical real-time vehicular applications, especially in urban mobility scenarios in which we focus on, since the total average execution time of all vehicles to estimate the positions meet the time constraint, which is around 100-300 ms. Furthermore, timing results can be more attractive  with smaller number of iterations $K$. Especially for a small network of 3-7 vehicles, methods converge within almost 1 iteration. Thus, the impact of a network delay in the overall time complexity may be smaller, realizing real-time applications. However, for a larger network of vehicles, much smaller $K$ is required (at the cost of lower accuracy) in order to meet time constraints.} Clearly, when VANET size increases, so does the average execution time of all three methods. Furthermore, \textbf{GLLMS} is proven to be faster than the other two schemes. Although both methods of \textbf{GLLME} and \textbf{GLCG} exhibit higher convergence speed and location accuracy due to the measurements exchanges step, they suffer from increased execution time, especially \textbf{GLLME}.

\begin{table}[htbp]
\caption{Total average execution time ($msec$)}
\begin{center}
\begin{tabular}{|c|c|c|c|c|}
\hline
\textbf{{K = 70 }} &
\textbf{\textit{\textbf{3 vehicles}}}&
\textbf{\textit{\textbf{7 vehicles}}}& \textbf{\textit{\textbf{10 vehicles}}}& \textbf{\textit{\textbf{15 vehicles}}}\\
\hline
$\textbf{GLLMS}$ & 
$\boldsymbol{8}$ &
$\boldsymbol{40}$ & $\boldsymbol{70}$ & $\boldsymbol{80}$\\
\hline
$\textbf{GLLME}$ & 
$\boldsymbol{20}$ &
$\boldsymbol{100}$ & $\boldsymbol{140}$ & $\boldsymbol{220}$ \\
\hline
$\textbf{GLCG}$ & 
$\boldsymbol{20}$ &
$\boldsymbol{70}$ & $\boldsymbol{80}$ & $\boldsymbol{130}$ \\
\hline
\multicolumn{5}{l}{}
\end{tabular}
\label{tab2}
\end{center}
\end{table}

\subsection{Experimental Evaluation on CARLA simulator}
\begin{figure}[htbp]
  \centering
  \subfloat[Ground truth]{\includegraphics[width=0.34\linewidth]{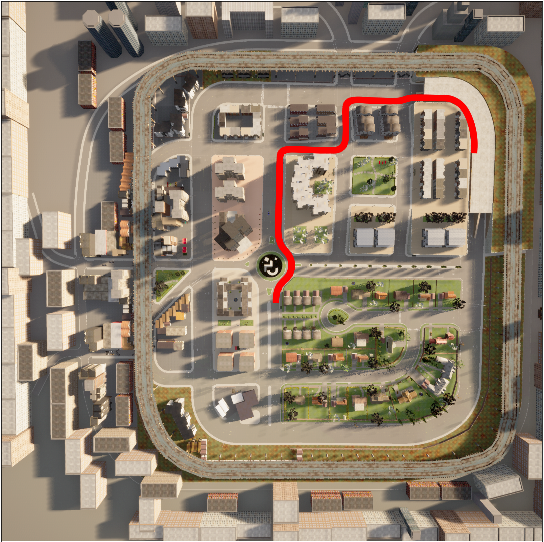}}
  \subfloat[Clusters]{\includegraphics[width=0.45\linewidth]{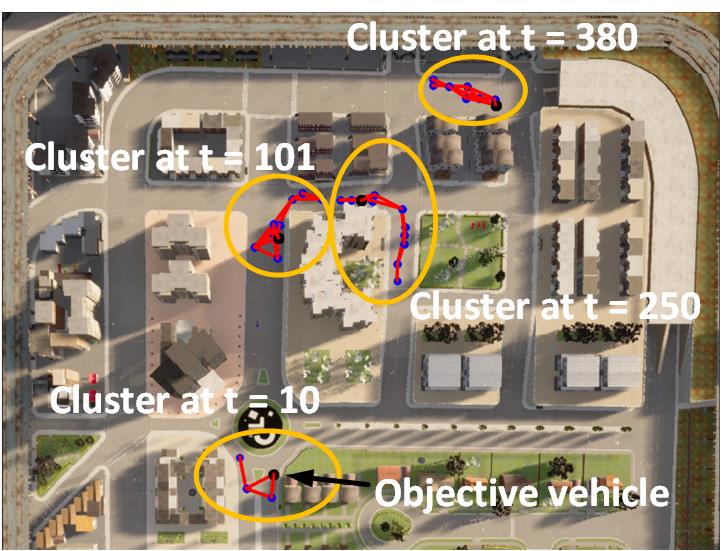}}
  \caption{Reference trajectory and clusters at four time instances in CARLA simulator}
  \label{traj-CARLA}
\end{figure}

The effectiveness of vehicular diffusion localization has been further validated using random realistic trajectories generated by CARLA. The latter is a renowned autonomous driving simulator, extensively used in various automotive applications, especially in computer vision based CAV perception. Therefore, we extracted the trajectory of a random objective vehicle (id 131) and those which belong to the same VANET or cluster as the objective, for $T = 448$ instances. We remind that clusters are formed by imposing a fixed communication range $r_c = 20m$ and maximum number of neighbors $N_{max} = 6$. Objective's trajectory and associated clusters at four distinct time instances are depicted in Fig.~\ref{traj-CARLA}. Black dot represent the objective vehicle. The size of clusters ranges between 2 and 26 vehicles. Total number of iterations is set to $K = 70$. Learning curves of AMSD have been omitted from evaluation, due to the fact that clusters don't contain the same vehicles as time evolves. Actually, different vehicles enter or exit the associated cluster during the simulation horizon. 

\begin{figure*}[htbp]
  \centering
  \subfloat[Localization Error of objective vehicle]{\includegraphics[width=0.3\linewidth]{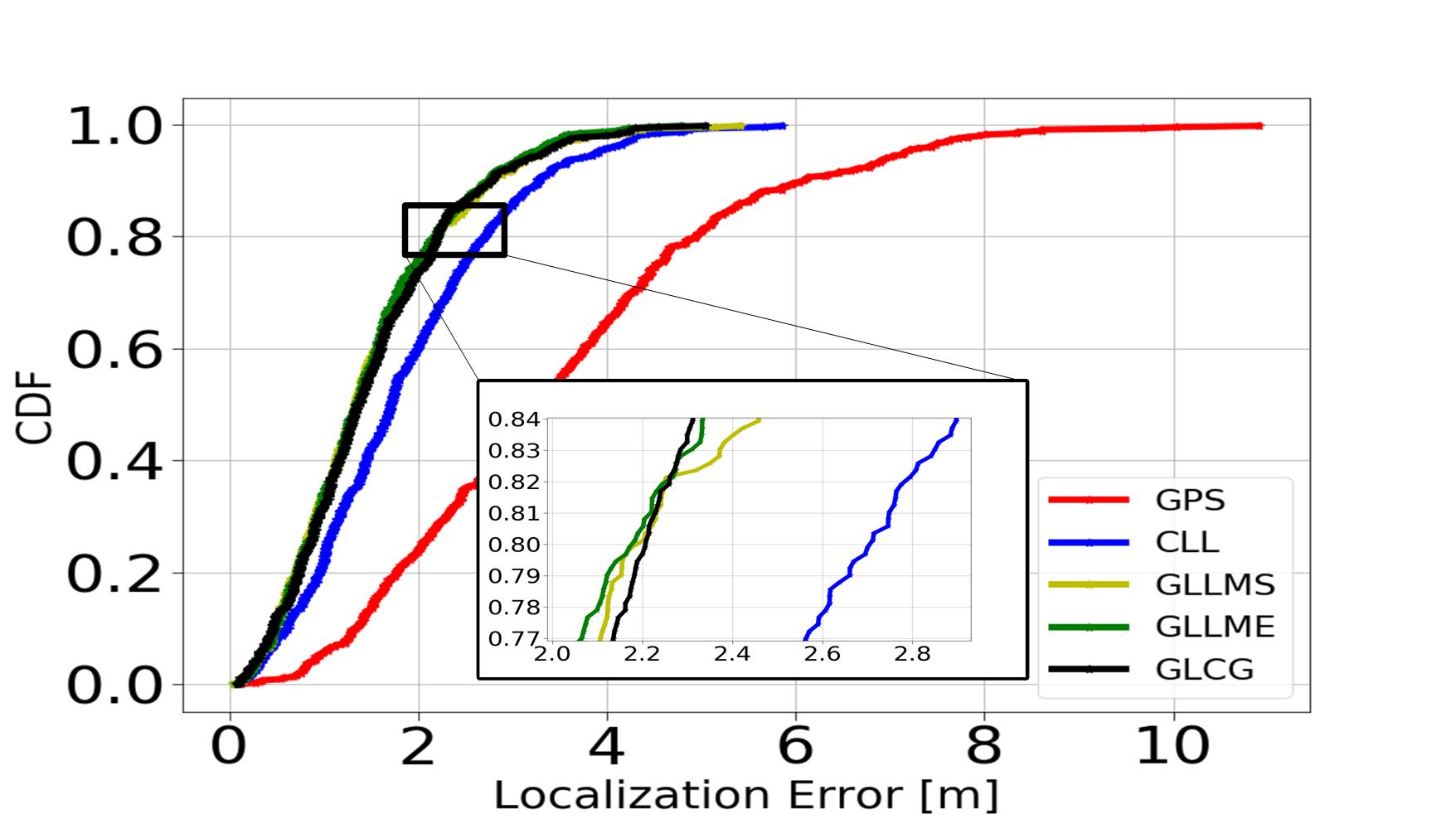}}
  \subfloat[Average Localization Error of cluster, \\ computed by objective vehicle]{\includegraphics[width=0.3\linewidth]{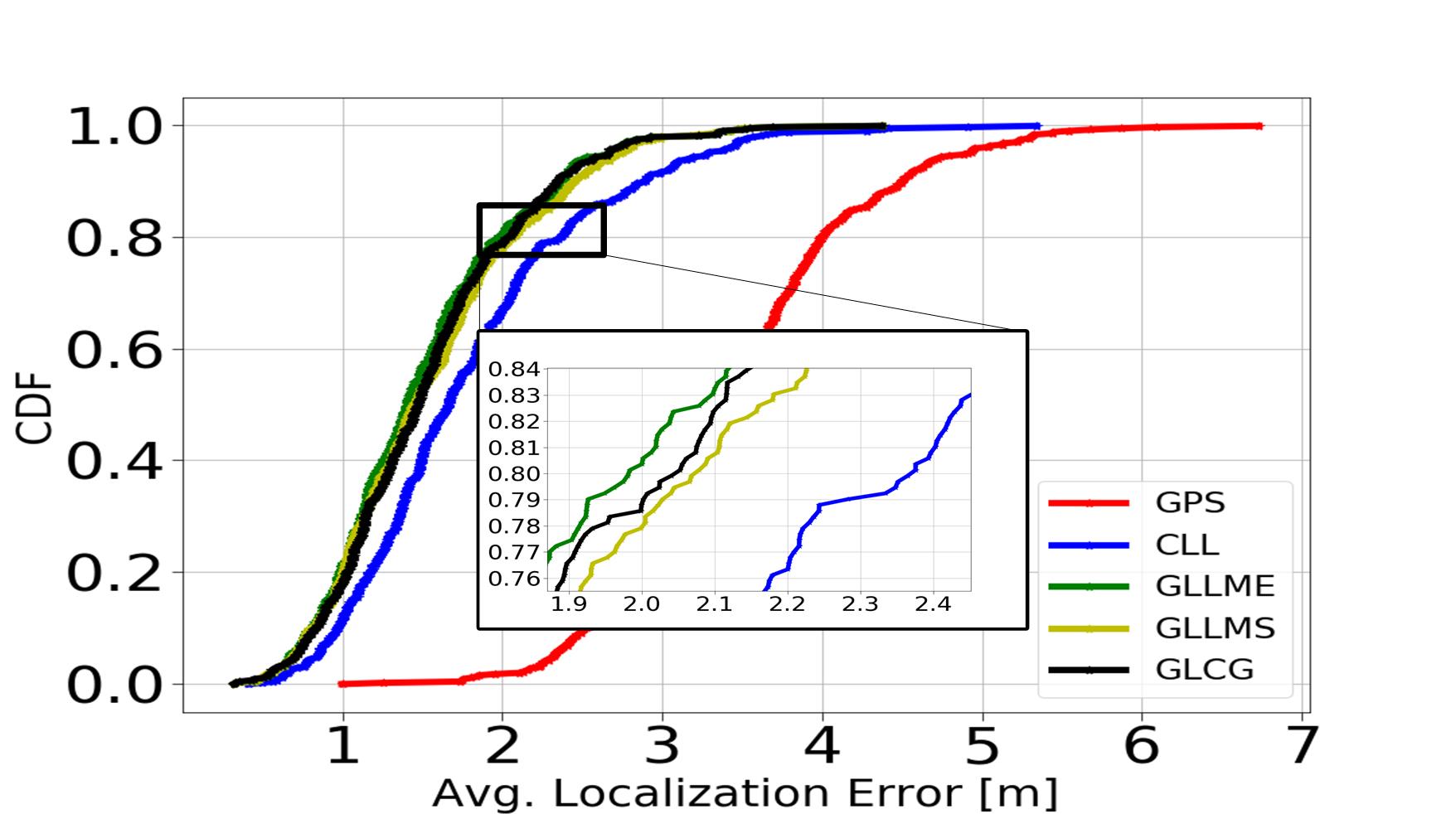}}
  \subfloat[Average Localization Error of objective vehicle]{\includegraphics[width=0.3\linewidth]{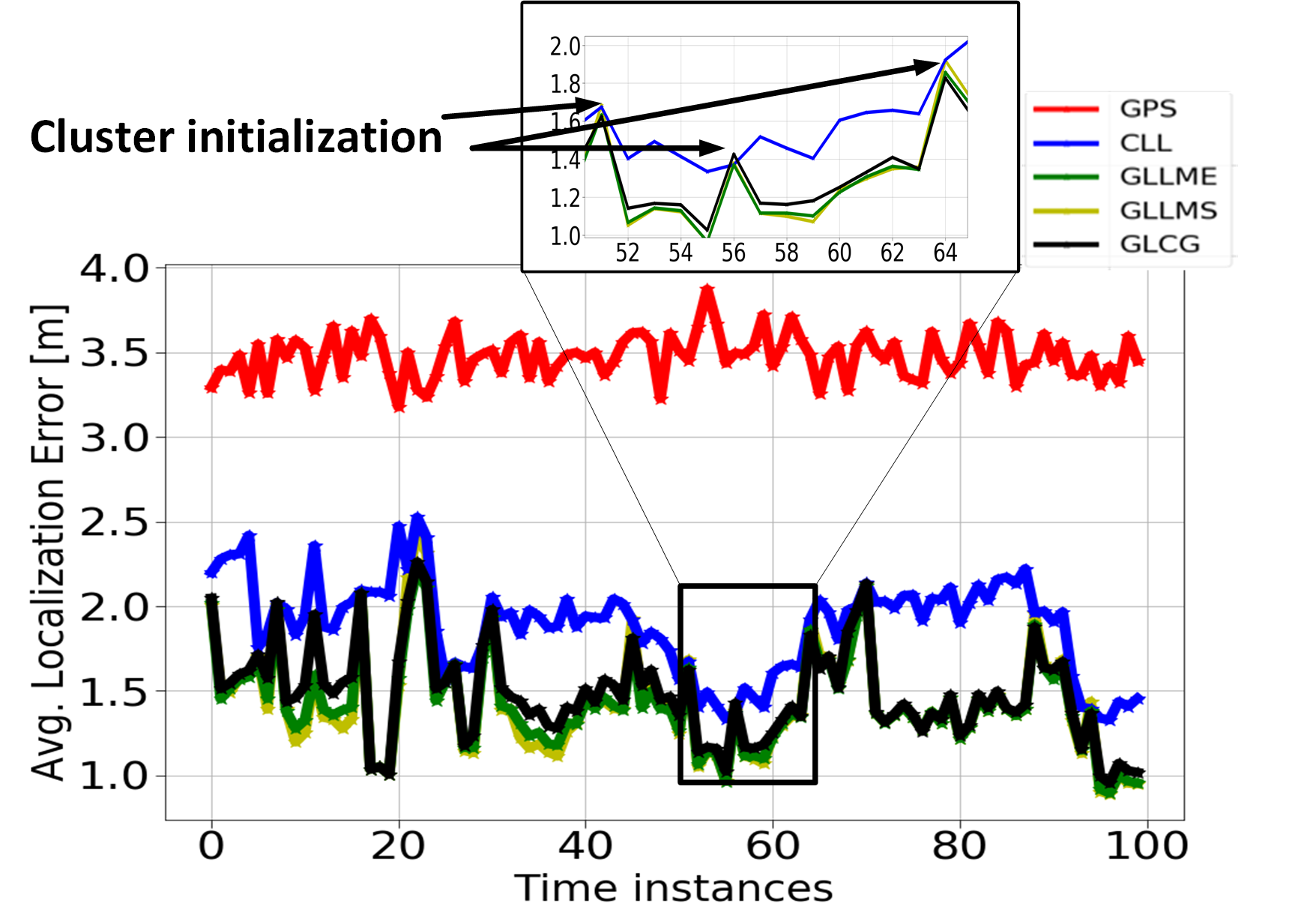}}
  \caption{Indicative statistical results in CARLA simulator}
  \label{CDF-ind}
\end{figure*}

In Fig.~\ref{CDF-ind}-(a), we demonstrate the Localization Error of objective vehicle, with $\sigma_d = 1m$ and $\sigma_{az} =4^\circ$. The reduction of GPS Localization Error is $\textbf{72\%}$ with $\textbf{GLLMS}$, $\textbf{GLLME}$ and $\textbf{GLCG}$ and 59\% with $\textbf{CLL}$. The overall location accuracy achieved by the objective vehicle, in terms of estimating the location vector of its cluster using the three proposed approaches, is depicted in Fig.~\ref{CDF-ind}-(b). We measured the Average Localization Error of the associated cluster using the entire location vector estimated by the objective vehicle each time instant. The reduction of GPS Average Localization Error is $\textbf{59\%}$ with $\textbf{GLLMS}$, $\textbf{60\%}$ with $\textbf{GLLME}$, $\textbf{59\%}$ with $\textbf{GLCG}$ and 51\% with $\textbf{CLL}$. Finally, in Fig.~\ref{CDF-ind}-(c) we plotted the Average Localization Error of objective vehicle over $200$ iterations, for the first $100$ time instances. Clearly, all three proposed approaches significantly outperformed $\textbf{CLL}$. Moreover, the peaks of diffusion towards the global solution of $\textbf{CLL}$ are due to the fact that at those time instances the associated cluster of objective vehicle is modified, since vehicles do constantly enter or exit. As matter of fact, instead of \textbf{Algorithm \ref{track}}, noisy GPS initialization is used during cluster initialization.

\begin{figure*}[htbp]
  \centering
  \subfloat[Objective, $t = 9$]{\includegraphics[width=0.3\linewidth]{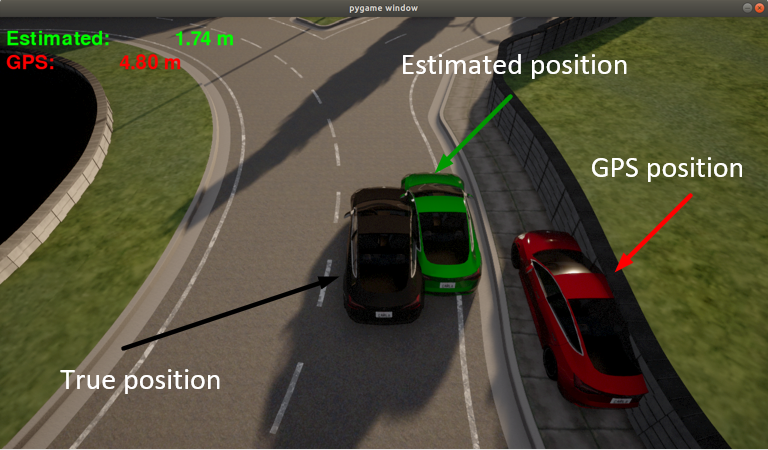}}
  \subfloat[Objective, $t = 18$]{\includegraphics[width=0.3\linewidth]{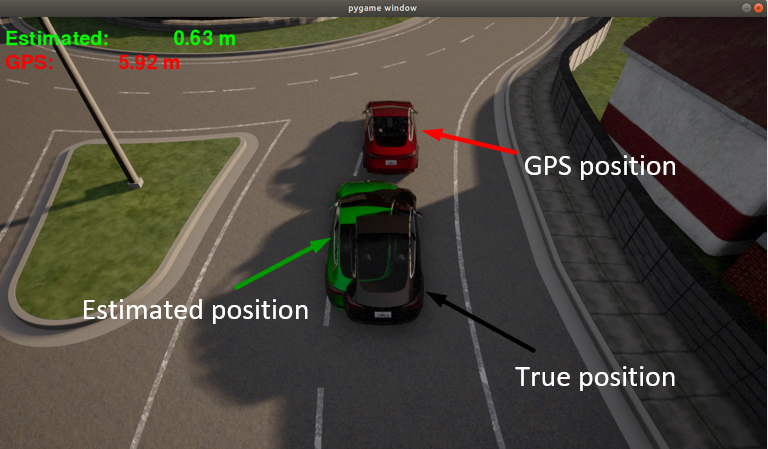}}
  \subfloat[Objective and neighbors location estimation, $t = 9$]{\includegraphics[width=0.3\linewidth]{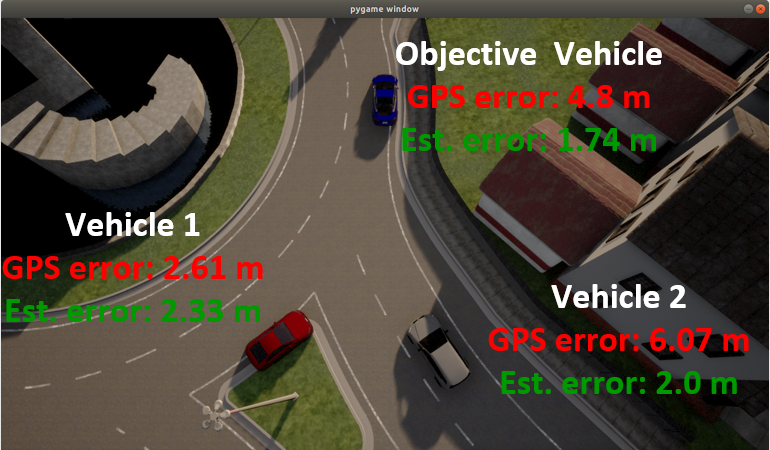}}
  \caption{CARLA visualization}
  \label{CARLA-vis}
\end{figure*}

Furthermore, we provide in Fig.~\ref{CARLA-vis} some indicative CARLA based results at time instances $t = 9$ and $18$, utilizing \textbf{GLLME}. In Fig.~\ref{CARLA-vis}-(a),(b), black vehicle corresponds to the true position, green one to the estimated by \textbf{GLLME}, while red vehicle is the actual GPS position. As you may see in both cases, objective's location accuracy is much higher than GPS ($1.74m$ vs $4.8m$ and $0.63m$ vs $5.92m$, respectively). The effectiveness of accurately estimating neighbor's location is also apparent in Fig.~\ref{CARLA-vis}-(c). At time instant $t=9$, objective vehicle and its two connected neighbors constitute the associated cluster. Obviously, \textbf{GLLME} estimates not only ego's ($1.74m$ vs $4.8m$) but also neighbor's location ($2.33m$ vs $2.61m$ and $2.0m$ vs $6.07m$) much more accurate than GPS. Consequently, all three proposed vehicular diffusion schemes achieved greater performance than GPS and global centralized solution of $\textbf{CLL}$, even in realistic urban traffic conditions generated by CARLA simulator. 

{
As a final remark from the conducted experiments, we have to point out that \textbf{GLCG} exhibits serious limitations and drawbacks as a cooperative localization approach, both in the presence of sensing (e.g., range measurements uncertainties) and communication-network failures, due to its highly vulnerability to noisy data. Hence, LMS based solutions seem to have the potential for efficient localization. The \textbf{GLLME} achieved higher convergence speed and accuracy, especially in the presence of a common network delay, while \textbf{GLLMS} exhibits low execution time and robustness to increased range measurements noise. The tradeoff between the two LMS methods is explicitly related to convergence and accuracy: an increased convergence speed of \textbf{GLLME} implies that a higher accuracy can be achieved earlier. Although the execution time of \textbf{GLLME} is higher, it is related to the required number of iterations $K$. For instance, if $K$ in the learning curve of Fig.\ref{vanet-size}-(c) is reduced almost by half, then \textbf{GLLME} converges (desired localization accuracy is attained) during $K$ iterations with total execution time almost $\sim 100$msec (from Table \ref{tab2}),  while \textbf{GLLMS} is far from the optimal solution. 
}

\section{Conclusion}
\label{conc}
In this paper, we treat the VANET as an undirected graph, encoding the V2V connections of operating vehicles via linear graph Laplacian operator. We have formulated three distributed and diffusion approaches for CAV localization, based on LMS, LMS with measurements exchanges and CG with measurements exchanges algorithms, utilizing the graph Laplacian operator to perform the multi-modal fusion. Each vehicle by interacting only with its local neighborhood, estimates the entire location vector of VANET in a fully distributed manner, adopting the ATC diffusion framework. All three methods, not only converged to the global \textbf{CLL} solution, but significantly outperformed it. Extensive simulations verify that both LMS and CG with measurements exchanges exhibited higher convergence speed with respect to single LMS. The latter proves its robustness under increased range measurements uncertainty. In all other experimental cases, LMS with measurements exchanges achieved the greatest performance in terms of convergence speed and location estimation accuracy. Due to the effectiveness of the proposed methods, we conclude that graph Laplacian distributed and diffusion localization will facilitate the design of much more efficient individual Path Planning and Control mechanisms. Future work will focus on MTL-based vehicular diffusion localization using realistic trajectories extracted by autonomous driving simulators, considering varying topologies of a large number of CAVs.

\section*{Acknowledgment}
We would like to thank Mr. Christos Anagnostopoulos for all of his support in extracting and visualizing data from CARLA autonomous driving simulator.

\ifCLASSOPTIONcaptionsoff
  \newpage
\fi



%
\bibliographystyle{IEEEtran}
\bibliography{Bib}

\section{Appendix}
\label{appen}
Covariance matrix $\boldsymbol{L_{i:}^{(t)T}}\boldsymbol{L_{i:}^{(t)}}$ is in fact a rank-one matrix, as a product of vector and its transpose. However, rank-one matrices have only one non-zero eigenvalue. Thus, $\lambda_1^{min} = 0$. Furthermore, the trace of a matrix is equal to the sum of its eigenvalues, i.e. to the largest eigenvalue in case of rank-one matrix. Therefore:
\begin{align*}
\lambda_1^{max} = tr(\boldsymbol{L_{i:}^{(t)T}}\boldsymbol{L_{i:}^{(t)}}) = \norm{\boldsymbol{L_{i:}^{(t)}}}^2= (\vert\mathcal{N}_i^{(t)}\vert-1)^2 + \vert\mathcal{N}_i^{(t)}\vert-1 .
\end{align*}

By the properties of positive semi-definite matrix (non-negative eigenvalues and at least one zero eigenvalue), we derive that $\lambda_2^{min} = 0$. As mentioned in \cite{Cattivelli2010a}, the largest eigenvalue of a real symmetric matrix is convex in the elements of that matrix. Thus, and due to the convexity of $c_{il}^{(t)}$, we have:
\begin{align*}
    \begin{split}
    \lambda_{max}\left(\sum_{l \in \mathcal{N}_i^{(t)}}c_{il}^{(t)}\left(\boldsymbol{L_{l:}^{(t)T}}\boldsymbol{L_{l:}^{(t)}}\right)\right) \leq \sum_{l \in \mathcal{N}_i^{(t)}}c_{il}^{(t)}\lambda_{max}\left(\boldsymbol{L_{l:}^{(t)T}}\boldsymbol{L_{l:}^{(t)}}\right)
    \end{split}
\end{align*}
\begin{align*}
    \begin{split}
    \leq \argmaxl_{l \in \mathcal{N}_i^{(t)}} \lambda_{max}\left(\boldsymbol{L_{l:}^{(t)T}}\boldsymbol{L_{l:}^{(t)}}\right) = \argmaxl_{l \in \mathcal{N}_i^{(t)}} \ tr(\boldsymbol{L_{l:}^{(t)T}}\boldsymbol{L_{l:}^{(t)}}) 	
    \end{split}
\end{align*}
\begin{align*}
    \begin{split}
    = \argmaxl_{l \in \mathcal{N}_i^{(t)}} \ \norm{\boldsymbol{L_{l:}^{(t)}}}^2 = \argmaxl_{l \in \mathcal{N}_i^{(t)}} \ ( (\vert\mathcal{N}_l^{(t)}\vert-1)^2 + \vert\mathcal{N}_l^{(t)}\vert-1)
    \end{split}
\end{align*}
\begin{align*}
    \begin{split}
    \Leftrightarrow \lambda_2^{max} \leq \argmaxl_{l \in \mathcal{N}_i^{(t)}} \ ( (\vert\mathcal{N}_l^{(t)}\vert-1)^2 + \vert\mathcal{N}_l^{(t)}\vert-1).
    \end{split}
\end{align*}

\begin{IEEEbiography}[{\includegraphics[width=1in,height=1.25in,clip,keepaspectratio]{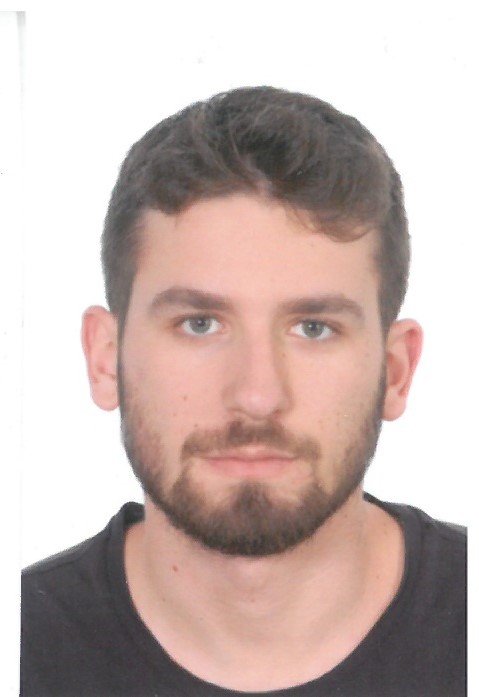}}]{Nikos Piperigkos}
received the Diploma and M.Sc. degrees from Computer Engineering and Informatics Department (CEID) of University of Patras, Greece in 2018 and 2020, respectively. Currently, he is a PhD student at the same department. Since 2017, he is a member of Signal Processing and Communications Lab of CEID. He joined the Industrial Systems Institute, "ATHENA RESEARCH CENTER" and became member of Multimedia Information Processing Group in 2019. He has participated in two Horizon 2020 R\&D projects. His research interests include cooperative localization and tracking, sensor fusion, intelligent transportation systems, distributed estimation, adaptive signal processing and learning algorithms.
\end{IEEEbiography}

\begin{IEEEbiography}[{\includegraphics[width=1in,height=1.25in,clip,keepaspectratio]{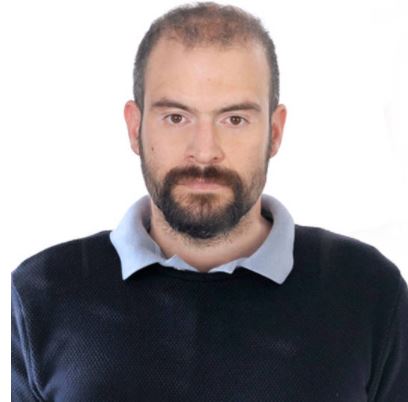}}]{Aris S. Lalos}
(Senior Member, IEEE) received
the Diploma, M.A.Sc., and the Ph.D. degrees
from the Computer Engineering and Informatics Department (CEID), School of Engineering
(SE), University of Patras (UoP), Rio Patras,
Greece in 2003, 2005, and 2010, respectively.
He has been a Research Fellow with Signal Processing and Communications Laboratory, CEID, SE, UoP, Rio-Patras, Greece, from
2005 to 2010, in Signal Theory and Communications (TSC) Department of the Technical
University of Catalonia (UPC), Barcelona, Spain, from October 2012–
December 2014 and in the Visualization and Virtual Reality Group from
January 2015 until the the present date. From October 2011 to October
2012, he was a Telecommunication Research Engineer with Analogies
S.A, an early stage start-up. In May 2018, Aris S. Lalos was elected
Principal Researcher (Associate Research Professor Level with tenure)
at Industrial Systems Institute, “ATHENA” Research Centre. He is an
author of 125 research papers in international journals (42), conferences
(78), and book chapters (5). His current research interests include, digital communications, adaptive filtering algorithms, geometry processing,
wireless body area networks, and biomedical signal processing.
Dr. Lalos has participated in more than 18 European projects related
to the ICT and eHealth domain  as a Project Coordinator (1), Technical Coordinator (1), WP Leader (8), Senior Researcher (15) and Researcher (3) and he is a Regular Reviewer for several
technical journals. He received the Best Demo Award in IEEE CAMAD
2014, Best Paper Award in IEEE ISSPIT 2015, World’s FIRST 10 K
Best Paper Award in IEEE ICME 2017 while, in January 2015, he
was nominated as Exemplary Reviewer for the IEEE COMMUNICATIONS
LETTERS.
\end{IEEEbiography}

\begin{IEEEbiography}[{\includegraphics[width=1in,height=1.25in,clip,keepaspectratio]{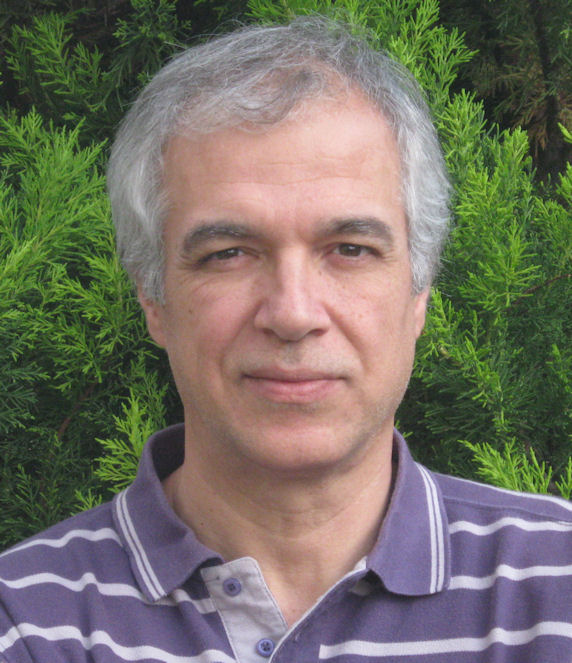}}]{Kostas Berberidis}
(S'87-M'90-SM'07) received the Diploma degree in electrical engineering from DUTH, Greece, in 1984, and the Ph.D. degree in signal processing and communications 
from the University of Patras, Greece, in 1990. During 1991, he worked at the Signal Processing Laboratory of the National Defense Research Center. From 1992 to 1994 and from 1996 to 1997,
 he was a researcher at the Computer Technology Institute (CTI), Patras, Greece. In period 1994/95 he was a Postdoctoral Fellow at CCETT/CNET, Rennes, France. 
Since December 1997, he has been with the Computer Engineering and Informatics Department (CEID), University of Patras, where he is currently a Professor,
and Head of the Signal Processing and Communications Laboratory.  Also, since 2008, he has been Director of the Signal Processing \& Communications Research Unit 
of the Computer Technology Institute and Press "Diophantus" and, since 2018, he is collaborating faculty with the Industrial Systems Institute, Athena Research Center, Greece.  
His research interests include distributed signal and information processing and learning, adaptive learning algorithms, signal processing for communications, 
wireless communications and sensor networks, array signal processing, smart grid, etc.

Prof. Berberidis has served or has been serving as member of scientific and organizing committees of several international conferences, 
as Associate Editor for the IEEE Transactions on Signal Processing and the IEEE Signal Processing Letters, 
as a Guest Editor for the EURASIP JASP and as Associate Editor for the EURASIP Journal on Advances in Signal Processing. 
Also, from February 2010 until December 2017 he served as Chair of the Greece Chapter of the IEEE Signal Processing Society.  
He has served in the "Signal Processing Theory and Methods” Technical Committee of the IEEE SPS, 
the “Signal Processing for Communications and Electronics” Technical Committee of the IEEE COMSOC 
and the EURASIP Special Area Team “Theoretical and Methodological Trends for Signal Processing”. 
Since January 2017 he serves as member of the Board of Directors (BoD) of EURASIP (second term started Jan. 2021).  
Moreover, since August 2015 he is a member of the EURASIP Technical Area Committee “Signal Processing for Multisensor Systems”. 
He is a member of the Technical Chamber of Greece, a member of EURASIP, and a Senior Member of the IEEE. 
\end{IEEEbiography}

\end{document}